\documentclass[oneside,11pt]{article}
\usepackage{amsmath,amsgen,amstext,amsbsy,amsopn,amsthm, amssymb}

\newtheorem{theorem}{Theorem}[section]
\newtheorem{lemma}[theorem]{Lemma}

\newtheorem{proposition}[theorem]{Proposition}

\theoremstyle{definition}
\newtheorem{definition}[theorem]{Definition}

\theoremstyle{remark}

\numberwithin{equation}{section}

\newcommand{\CA}{{\mathcal A}}

\newcommand{\CF}{{\mathcal F}}
\newcommand{\CG}{{\mathcal G}}
\newcommand{\CH}{{\mathcal H}}
\newcommand{\CK}{{\mathcal K}}
\newcommand{\CL}{{\mathcal L}}
\newcommand{\CM}{{\mathcal M}}
\newcommand{\CP}{{\mathcal P}}
\newcommand{\CW}{{\mathcal W}}
\newcommand{\I}{{\rm i}}
\newcommand{\E}{{\rm e}}

\newcommand{\Imt}{\Im m\,}
\newcommand{\ad}{{\rm Ad}\,}

\newcommand{\norm}[1]{\|#1\|}
\newcommand{\BR}{\mathbb{R}}
\newcommand{\BC}{\mathbb{C}}
\newcommand{\sabsatz}{\par\smallskip\noindent}
\newcommand{\mabsatz}{\par\medskip\noindent}

 \def\mfrac#1#2{\hbox{$\frac{{#1} }{ {#2}}$}}
\begin{document}
\title{\bf{On the PCT--Theorem\\ in the Theory of 
 Local Observables}}
 \author{\vspace{5pt} H.J. Borchers$^{1}$ and J. Yngvason$^{2}$\\ 
 \vspace{-4pt}\small{$1.$ The Erwin Schr\"odinger International
Institute for Mathematical Physics} \\ \small{Boltzmanngasse 9, A--1090 Wien}\\
\small{and}\\ \small{Institut 
f\"ur Theoretische Physik,
Universit\"at G\"ottingen,}\\ \small{Bunsenstrasse 9, D--37073 
G\"ottingen\thanks{permanent address}}\\
\vspace{-4pt}\small{$2.$ Institut f\"ur Theoretische Physik,
Universit\"at Wien}\\ \small{Boltzmanngasse 5, A 1090 Vienna,
Austria}\\ \\ \small{ \it Dedicated to Sergio Doplicher and John Roberts
on the occasion of their 60th birthday}}
\date{}
 \maketitle

\begin{abstract}
We review the PCT-theorem and problems connected with its demonstration.
We add a new proof of the PCT-theorem in the theory of local observables
which is similar to that one of Jost in Wightman quantum field theory.
We also look at consequences in case the PCT-symmetry is given on the
algebraic level.  At the end we present some examples which 
answer general questions and throw some light on open problems.
\end{abstract}

\maketitle

\section{Introduction}
Before starting to discuss the different aspects of the PCT-theorem
we collect some terminology and notations that we shall be using. 
This is necessary because
the notions in the literature are not unique.
\subsection{\bf On quantum field theory}
The term Lagrangian or Wightman field theory means quantum field
theory with \lq\lq point like\rq\rq\ 
localized fields.  More precisely, the fields
are operator valued distributions and the field operators are supposed
to transform covariantly under a continuous unitary representation of
the connected part $\CP^\uparrow_+$ of the Poincar\'e group, or at
least of the subgroup of space-time translations.  The representation
of the translations is required to fulfill the spectrum condition and
the representation space shall contain a vacuum vector (mostly assumed
to be unique).  It is also assumed that this vacuum vector is cyclic
for the algebra generated by the field operators.  For space-like
separation of the arguments the field operators shall either commute
or anti-commute.  The Bose-fields have to commute with themselves and
all other fields.  Fermi-fields anti-commute with Fermi-fields.  The
Bose-fields represent particles with integer spin and the Fermi-fields
such with half-integer spin.  Therefore, the phrase ``spin and
statistics'' means in this context the same as ``spin and commutation
relations''.  Para-fields, introduced by O.W. Greenberg \cite{Gr61},
will not be considered.  Para-statistics are usually only discussed in
the frame of quantum field theory of local observables, where a
reduction to ordinary Bose- or Fermi commutation relations has 
been achieved by Doplicher and Roberts in \cite{DR90}.

Quantum field theory of local observables (QFTLO) in the sense of
Araki, Haag and Kastler \cite{Ha92} is concerned with $C^*$-algebras
$\mathcal{A}(O)$, associated with bounded open regions $O\subset
\mathbb R^d$.  These algebras shall fulfill isotony, i.e., $O_1\subset
O_2$ implies $\CA(O_1)\subset\CA(O_2)$, and locality, i.e., if $O_1$
and $O_2$ are space-like separated, then the algebras $\CA(O_1)$ and
$\CA(O_2)$ commute element-wise.  Space-like separation is here
defined in terms of the Minkowski metric on $\mathbb R^d$.  For
unbounded domains $G$ the algebra $\CA(G)$ is defined as the
$C^*$-inductive limit of $\{\CA(O):\ O\subset G\}$.  
Usually it is
assumed that the net $\{\CA(O)\}$ is covariant under a representation
of the proper, orthochronous Poincar\'e group $\CP^\uparrow_+$, or at 
least some subgroup $\CG\subset\CP^\uparrow_+$,  by 
automorphism of
$\CA(\BR^d)$, i.e., to every $g\in\CG$ there is assigned an
automorphism $\alpha_g$ such that
$\alpha_{g_{1}}\alpha_{g_{2}}=\alpha_{g_{1}g_{2}}$, and for all $O$
\begin{equation}\label{1.0}\alpha_g\CA(O)=\CA(gO).\end{equation}
    
A QFTLO may often be defined in terms of Wightman fields by taking as 
local algebras $\CA(O)$ the von Neumann algebras generated by the polar 
decompositions of the smeared field operators with test functions 
supported in $O$. The local commutativity of these algebras is 
not a consequence of the locality of the field alone, but several
conditions which guarantee this are known. See \cite{BY92} for a 
review.

\subsection{\bf Representations.}
A representation $\pi$ of $\CA(\BR^d)$ on a Hilbert space $\CH_{\pi}$
will always mean a non-degenerate representation.  We shall denote the
von Neumann algebra $\pi(\CA(O))^{\prime\prime}$ by $\CM_{\pi}(O)$, 
or simply by $\CM(O)$ if $\pi$ is a vacuum representation (see below).
We say that {\it weak additivity} holds in the representations 
considered, i.e., the von Neumann algebra generated by 
$\cup\{\CM_{\pi}(O+x):
x\in\BR^d\}$ is equal $\CM_{\pi}(\BR^d)$ for each bounded, open $O$, 
independent of its size. Usually it is required only for vacuum
representations.
When the QFTLO is
covariant under $\CP^\uparrow_+$ or some subgroup $\CG\subset
\CP^\uparrow_+$ the automorphisms $\alpha_{g}$ are {\it implemented} in the
representation $\pi$ if there is a continuous unitary representation $U$ of
$\CG$ on $\CH_{\pi}$ such that
\begin{equation}\label{1.1}U(g)\pi(A)U^*(g)=\pi(\alpha_gA),\qquad
A\in\CA(\BR^d),\ g\in\CG.
\end{equation}
By \eqref{1.0}, it is clear that 
\begin{equation}\label{1.1a}U(g)\CM_{\pi}(O)U^*(g)=\CM_{\pi}(gO).
\end{equation}

For a classification of the representations $\pi$ one often takes
for $\CG$ the translation group $\BR^d$ or the subgroup of time
translations.
If a topological group $\CG$ acts on a $C^*$-algebra $\CA$ 
by automorphisms $\alpha_g$ one speaks of a {\it $C^*$ dynamical system}.
If a representation $\pi$ of $\CA$ is given, 
then it is of interest to know whether or not the group  action  
is unitarily implemented in the representation. The answer is known up to 
multiplicity
problems implying that eventually one might have to to change the
multiplicity of the representation. Representations which differ
at most in their multiplicity are called {\it quasi-equivalent}.
Such representations have the same {\it folium} of states, i.e., states 
given by a density matrix in the representation. These are the
ultra-weakly continuous (normal) states on the von Neumann algebra 
$\pi(\CA)^{\prime\prime}$. 
For a discussion of the problem of implemented group 
action and its history see \cite{Bch96},
Section II.7. The main result is:
 
\begin{theorem} {\rm (Borchers).}
Let $\{\CA,\CG,\alpha\}$ be a $C^*$-dynamical system, with $\CG$ a
topological group, and let $\pi$ be a representation of $\CA$.
The following are equivalent:
\begin{itemize}
\item[(i)]
There exists a representation $\widehat{\pi}$ of $\CA$ that is 
quasi-equivalent to $\pi$
and such that $\alpha_g$ is implemented in the 
representation
$\widehat{\pi}$ by a continuous unitary
representation of $\CG$.
\item[(ii)]
The folium $F(\pi)$ of $\pi$ is invariant
under the adjoint action $\alpha_g^*$ and the action is strongly
continuous in $g$ on $F(\pi)$ .
\end{itemize}
\end{theorem}

A {\it thermal representation} is characterized by the group of time
translations. Its representation $U(t)$ must have an invariant vector
$\Omega_\beta\in\CH_\pi$ which is cyclic and separating for the representation.
Moreover, the representation $U(t)$ shall fulfill the $\beta$-KMS condition,
which will be explained together with the Tomita-Takesaki theory.

We call a 
representation $\pi$ a {\it particle representation}
if the whole translation group $\BR^d$ is implemented by a 
unitary group $U(a)$ that fulfills the spectrum condition, i.e.,
the spectrum of $U$ is contained in the closed, forward light-cone 
$\overline V^+$. 
The name ``particle representation'' will be used in accordance with the
terminology in \cite{Bch96}.
This does not imply that the mass operator has a discrete part. 
In nature probably most massive particles are infra-particles implying
that there are representations such that the mass operator has a 
purely continuous spectrum.
A particle representation $\pi$ is 
called a {\it vacuum representation}, if $U(a)$ has an invariant vector
$\Omega$ and if this vector is cyclic for the representation.
Moreover, we will assume that vacuum representations are factor 
representations. A vacuum representation will always be denoted by $\pi_0$.

If $\pi$ is a representation with cyclic vector $\Omega$ then we say 
that the {\it Reeh-Schlieder property\/} holds for
$(\pi,\Omega)$, if $\Omega$ is
cyclic for every $\CM_{\pi}(O)$ with $O$ open and nonempty. If $\pi$ 
is a particle representation enjoying the weak additivity property
and if $\psi\in\CH_\pi$ has compact energy support then
$(\pi,\psi)$ has the Reeh-Schlieder property.  For the
vacuum vector it  was proved by Reeh and Schlieder \cite{RS61}

\subsection{\bf Maximal local algebras.}
Domains of special interest in our discussion are the {\it wedges} in
$\BR^d$.  They are defined in the following manner 
by two non-zero light-like vectors
$\ell_1,\ell_2$, belonging to the boundary of the forward light-cone
$V^+$ :
\begin{equation}W(\ell_1,\ell_2)=\{\lambda\ell_1+\mu\ell_2+\ell^\perp:
\lambda>0, \mu<0,(\ell^\perp,\ell_i)=0, i=1,2\}.  \end{equation} The
plane spanned by $\ell_1$ and $\ell_2$ will be called the {\it
characteristic two-plane} of the wedge $W(\ell_1,\ell_2)$.  The
translated wedge $W(\ell_1,\ell_2)+a$ is denoted by $W(\ell_1,\ell_2,a)$.
The $C^*$-algebra $\CA(W)$ is the algebra generated by all
$\CA(O)$ with $O\subset W$.  A representation $\pi$ of the theory of local
observables fulfills {\it wedge duality}, if for every wedge $W$ one has
\begin{equation}\CM_{\pi}(W)'=\CM_{\pi}(W'),\end{equation}
where $W'$ denotes the interior of the space-like complement of $W$, i.e.
$W(\ell_1,\ell_2)'=W(\ell_2,\ell_1)$
and $\CM_\pi(G)$ the von Neumann algebra generated by $\pi(\CA(G))$.
The index $\pi$ will be dropped if $\pi$ is a vacuum representation.

Another important family of domains are the double cones or other
domains which are the open interior of intersections of wedges. 
Double cones, denoted by $D$, play a special role. At several occasions
one assumes or derives some properties of the algebras
$\CM_{\pi}(W)$ associated with wedges and one wants to deduce
corresponding properties of a double cone algebra. This is only 
possible if the latter algebra can be expressed in terms of wedge algebras.
(For short we often use the term ``double cone algebra'' instead of 
``algebra associated
with a double cone'', and analogously for other domains.) 
Namely, the double cone
algebra must have the form
\begin{equation}\label{maximal}\CM_{\pi}(D)=\cap\{\CM_{\pi}(W):\ D\subset W\}.
\end{equation}
If this is the case then we call $\CM_\pi(D)$ a {\it maximal
local algebra}. It is easily checked that the theory constructed from the 
maximal local algebras fulfills all requirements of the QFTLO. If
$\CM_{\pi}(D)$ is not the maximal local algebra then one can {\it 
define}
$\CM_{\pi,{\rm max}}(D)$ by the right hand side of Eq. \eqref{maximal}.
If wedge duality holds, these algebras are, indeed, the maximal 
extensions of $\CM_{\pi}(D)$ compatible with locality.
In general the QFTLO's $\{\CA(O)\}$ and $\{\CM_{\pi,{\rm max}}(O)\}$
will have different families of representations. We shall
consider the maximal algebras only for a vacuum representation and denote
them in this case simply by
$\CM_{{\rm max}}(O)$.

For reasons we will see later, several aspects of QFTLO can be handled
by taking as input von Neumann algebras $\CM(W)$ associated only with wedges. 
Of course the algebras associated with $W$ and $W'$ have to commute. 
In this case one can always define double cone algebras as on the 
right hand side of \eqref{maximal} and these algebras will satisfy 
locality. The only problem is that in general one does not know the 
their size, in particular $\CM(W)$ need not be generated by the 
$\CM(D)$'s with $D\subset W$.

\subsection{\bf Charge sectors.}
Two representations $\pi_1,\pi_2$ of $\CA(\BR^d)$ are quasi-equivalent 
if they have the same kernel and if the isomorphism
$\pi_1(\CA(\BR^d))\leftrightarrow \pi_2(\CA(\BR^d))$ extends to an
isomorphism between the von Neumann algebras generated by the two
representations.  
For a $C^*$-algebra as
complicated as (non-trivial) QFTLO there exists at least a continuum
of non-equivalent representations.  Therefore, one is interested in
principles which group the set of representations into sub-families. 
The principle mostly used is that of {\it local equivalence}.  Two
representations $\pi_1,\pi_2$ are called {\it locally equivalent}, if
for every bounded open region the representations $\pi_1(\CA(O))$ and
$\pi_2(\CA(O))$ are quasi-equivalent.  If one of them is a vacuum
representation then one calls the second representation {\it locally
normal}.  The requirement of local normality is often used in order to
select from the set of thermal representations (defined by other
means) a suitable sub-family which, from some point of view, can be
regarded as physically acceptable.

While in the Lagrangian or in Wightman's field theory charged fields
are put in by hand, it is the philosophy of the QFTLO that
representations of the observable algebra describing a finite
number of charged particles shall be constructed from the algebra of
observables.  Also the charged fields connecting the vacuum
representation with the representations describing charges should be
constructed with help of the different representations.  This can be
worked out if one uses the equivalence relation introduced by Borchers
\cite{Bch65}. If $\pi_0$ is a vacuum representation, then a factor
representation $\pi_1$ is called a {\it charged sector} 
if $\pi_0$ and $\pi_1$ have the
same kernel and if for every $O$ the isomorphism $\pi_0(\CA(O'))
\leftrightarrow \pi_1(\CA(O'))$ extends to an isomorphism of the
corresponding von Neumann algebras.  Here $O'$ denotes the space-like
complement of $O$.  With this concept Doplicher, Haag and Roberts
\cite{DHR69a,DHR69b}, \cite{DR90} have worked out the details of the mentioned program. 
The algebra generated by the local observables and by the localized
charged fields is called the {\it field algebra} and the corresponding net 
is usually denoted by
$\{\CF(O)\}$. Within this setting also the concept of {\it 
conjugate charge sectors} has a precise meaning: If an element $F$ of the 
field algebra generates a charged sector by applying it to the vacuum, 
then the conjugate sector is generated in the same way by $F^{*}$. 

\subsection{\bf Gauge transformations.}
We shall use the term {\it gauge transformation} for any unitary 
operator $U$ on the Hilbert space of a vacuum representation 
$\pi_{0}$ of a QFTLO that fulfills $U\Omega=\Omega$ and
\begin{equation} U\pi_0(\CA(O))U^*=\pi_0(\CA(O))\end{equation}
for every bounded open $O$.  In Lagrangian field theory such operators
typically implement transformations of the form $\psi(x)\to
\E^{\I\varphi}\psi(x)$ and generate a compact group.  Since the QFTLO
is given by a set of axioms and not by a finite number of fields the
gauge group in the sense defined is not compact in general, however. 
In the last section an example of such a case will be considered. 
Therefore, a possible requirement for selecting a reasonable family of
QFTLO is the assumption that the gauge group is compact.  
There are
other conditions implying this property.  Buchholz and Wichmann
\cite{BuW86} have introduced the concept of {\it nuclearity}.  This is the
requirement that the number of states, which can be created locally,
does not increase too fast with energy.  Doplicher and Longo
\cite{DL84} have introduced the {\it split property}.  This is equivalent to
the following: Let $O_1\subset O_2$ be two bounded open regions such
that the closure of $O_1$ is still contained in $O_2$.  Let $\omega_1$
be a vector state on $\pi_0(\CA(O_1))$ and $\omega_2$ be a vector
state on $\pi_0(\CA(O'_2))$.  Then exists a vector state on $\pi_0$ of
the QFTLO which coincides with $\omega_1$ on $\pi_0(\CA(O_1))$ and
with $\omega_2$ on $\pi_0(\CA(O'_2))$.  Between these concepts one has
the following relations:
\begin{equation}
\text{nuclearity property}\longrightarrow
\text{split  property} \longrightarrow
\text{compact gauge group}
\end{equation}
For the first arrow see \cite{BuW86} and
\cite{BDF87}, for the second see \cite{DL84}.

\subsection{\bf Tomita-Takesaki theory.}
The main tool for handling the problems connected with the PCT-theorem
in the QFTLO is the modular theory introduced by Tomita. It is usually
called Tomita-Takesaki theory because the first presentation of this 
theory, beyond a preprint, is due to Takesaki.

Let $\CH$ be a Hilbert space and $\CM$ be a von Neumann algebra,
acting on this space, with commutant $\CM'$. A vector $\Omega$ is cyclic
and separating for $\CM$, if $\CM\Omega$ and $\CM'\Omega$ are dense in $\CH$.
If these conditions are fulfilled, then a modular operator $\Delta$
and a modular conjugation $J$ are associated with the pair $(\CM,\Omega)$,
such that
\begin{itemize}
\item[(i)] $\Delta$ is self-adjoint, positive and invertible,
\[
\Delta\Omega=\Omega,\qquad J\Omega=\Omega.
\]
\item[(ii)] The operator $J$ is a conjugation, i.e., $J$ is anti-linear,
$J^*=J$, $J^2=1$, and $J$ commutes with $\Delta^{\I t}$. This implies
the relation 
\[
\ad J\Delta =\Delta^{-1}.
\]
\item[(iii)] For every $A\in\CM$ the vector $A\Omega$ belongs to the domain
of $\Delta^{\frac12}$, and
\[
J\Delta^{\frac12}A\Omega=A^*\Omega=:SA\Omega.
\]
\item[(iv)] The unitary group $\Delta^{\I t}$ defines a group of automorphisms
of $\CM$,
\[
\ad \Delta^{\I t}\CM =\CM, \qquad \text{for all } t\in\BR.\]
\item[(5)] $J$ maps $\CM$ onto its commutant
\[
\ad J\CM  =\CM'.\]
\end{itemize}

These results apply in particular to QFTLO in a vacuum 
representation because of the Reeh-Schlieder theorem which implies that 
the vacuum vector is cyclic and separating for every algebra
$\CM(G)$, where $G$ is any domain which has a space-like complement
with interior points.

The matrix elements of the 
modular group have the following important analyticity properties. 
If $A,B\in\CM$ and $\sigma^t(A):
=\ad\Delta^{\I t} A$ then the continuous function
\[
F(t)=(\Omega,B\sigma^t(A)\Omega)
\]
has bounded analytic continuation into the strip $S(-1,0)=\{z\in\BC:
-1<\Imt z<0\}$. At the lower boundary one finds
\begin{equation}F(t-\I)=(\Omega,\sigma^t(A)B\Omega)\end{equation}
This relation is called the {\it KMS-condition}. In QFTLO the
group $\sigma^t$ 
coincides for thermal states with scaled time translations. If  
$\tau$ denotes the automorphism of the time translations and $\Omega$ 
defines a thermal state at inverse temperature $\beta$ the
equations become
\begin{equation}\label{betakms}
\begin{split}
F(t)&=(\Omega,B\tau_t(A)\Omega),\\
F(t-\I\beta)&=(\Omega,\tau_t(A)B\Omega).
\end{split}
\end{equation}
 The relations
Eq.\ \eqref{betakms} are called the {\it $\beta$-KMS-condition}.

\section{Review of the PCT-theorem}

\subsection{Point-fields}
In Lagrangian or Wightman field theory the PCT-ope\-ra\-tor $\Theta$
is an anti-unitary operator implementing the PCT-symmetry (if present).
For a scalar field one has the relation
\begin{equation}\Theta\Phi(x)\Theta=\Phi^*(-x).\end{equation}
The formula for higher spin looks the same with respect to the
space-time variable $x$, but in the index space one has in general to introduce
an additional transformation independent of the variable $x$. In 
a suitable spinorial basis the matrix of this transformation is 
diagonal and each component simply gets multiplied by a phase factor. 
In addition $\Theta$ shall fulfill 
the following commutation rule with the Poincar\'e transformations
\begin{equation}\label{thetapoinc}\Theta U(\Lambda,a)\Theta=U(\Lambda,-a).\end{equation}
These requirements imply
\begin{equation}\Theta=\Theta^{-1}=\Theta^*.\end{equation}
PCT-theorems are results showing the existence of a PCT-operator 
$\Theta$. The product PT represents an element of the Lorentz
group. All proofs so far assume that PT is the element $-1$
and has the determinant $+1$.
This implies that the Minkowski space must have even dimensions. In the 
odd dimensional case one replaces P by the total
reflection in the space perpendicular to the one-direction, denoted 
P$_1$. We will not
discuss this case and restrict ourselves to even dimensional Minkowski
spaces. As remarked in the last section, however, also in odd 
dimensions, where PT has determinant $-1$, free single component 
hermitian fields have full PCT symmetry, even 
without Lorentz covariance. 

The Lagrange function for a field theory is usually invariant under
total reflection, time-reversal, and charge conjugation. But it does not
necessarily mean that these symmetries are implemented separately by 
unitary or antiunitary operators.
G. L\"uders \cite{Lue54} discovered that the product of these symmetries
is always implemented by an operator $\Theta$, the PCT-operator. The
input for this result is the implemented Poincar\'e covariance,
spectrum condition, and the existence of charge conjugate partners.
Pauli \cite{Pau55} was quite excited about this result, because it clari\-fied
the relation about spin and the commutation relations of the fields.
It also gave an understanding why for Fermi-fields one should use 
a positive energy representation and not one with a "Dirac sea".
(For the discussion of spin and statistics see also G. L\"uders and
B. Zumino \cite{LZ58}).

In Wightman's field theory \cite{Wi56} the PCT-theorem was proved
by R. Jost \cite{Jo57}. This proof is based on the same assumptions as for the 
Lagrangian field theory which are
\begin{itemize}
\item[{1.}] The theory is Poincar\'e covariant and the translations
fulfill the spectrum condition. Moreover, the representation space contains
a cyclic vacuum vector.
\item[{2.}] The Poincar\'e transformations of the fields induce
a finite dimensional representation of the Lorentz group in the index space.  
\item[{3.}] Every field in the theory is accompanied by its charge conjugate
partner (which may be the field itself). 
\end{itemize}
Jost's proof is based on a result of Hall and Wightman \cite{HW57}
which says that the analytic continuation of the Wightman functions
are invariant under the complexified connected component of the
Lorentz group.  In even dimension the complex Lorentz group contains
the element $-1$.  The requirement 2 does not forbid that a certain
field is infinitely degenerate.  But if we deal with fields
transforming with an irreducible, infinite dimensional representation of the Lorentz group
in the index space the properties leading to the spin-statistics and PCT-theorem may be
destroyed.  R. Streater \cite{Str67} constructed examples where fields
with integer spin anti-commute and those with half-integer spin
commuted.  Oksak and Todorov \cite{OT68} gave other examples where
Jost's proof of the PCT theorem did not work. Requirement 3 implies
a symmetry under complex conjugation for the set of Wightman functions.

\subsection{\bf Algebraic PCT-symmetry in QFTLO}
Before discussing the PCT-theorem in the QFTLO let us assume that we have 
such symmetry on the algebraic level, i.e., there exists an anti-linear
automorphism $\theta$ of $\CA(\BR^d)$ with
\begin{equation}\label{theta}
\begin{split}
\theta(AB)&=\theta(A)\theta(B),\qquad \theta(\lambda A)=
\overline{\lambda}\theta(A),\\
\theta(\CA(O))&=\CA(-O),\qquad A,B\in\CA(\BR^d).\end{split}\end{equation}

The automorphism $\theta$ should 
also transform the translations in the correct manner, i.e., 
\begin{equation}\label{thetaalpha}\theta(\alpha_a(A))=\alpha_{-a}(\theta(A)).\end{equation} 
Since $\theta$ maps the algebra $\CA(\BR^d)$ onto itself, its 
transpose,
$\theta^*$, maps the dual space $\CA(\BR^d)^*$ onto itself. This implies
in particular that $\theta$ is represented by an anti-unitary operator
on the standard representation of the enveloping von Neumann algebra
$\CA(\BR^d)^{**}$. For later discussions it is of interest to know
sub-families of representations which are mapped by the PCT-symmetry
onto itself.

\begin{theorem} 
Let $\{\CA(O)\}$ be a QFTLO with algebraic PCT-symmetry $\theta$. Assume in addition
that the translations $\alpha_a$ act strongly continuous, i.e., the
function $a\to\alpha_a(A)$ is continuous in the norm topology of
$\CA(\BR^d)$ and this for every $A\in\CA(\BR^d)$. Then the family
of particle representations is invariant under $\theta$.
Moreover, the sub-family of vacuum representations is also 
invariant.\end{theorem}

The proof will be given in the next section. The requirement that the
translations act strongly continuous is not necessary. For the proof of 
the general case one would have to go into details of positive
energy representations as described in \cite{Bch96}.
But since we will look only at representations which are locally normal,
the general version of Thm. (2.1) is not needed for the following
reason: 
Let $\{\CA(O)\}$ be a QFTLO and $\pi_0$ a vacuum representation.
Then $\{\CM_{\pi_{0}}(O)\}$ defines a new QFTLO, and 
the locally normal representations of
the two nets are clearly in one to one correspondence with each other. 
Now let $\pi_{1}$ be a locally normal representation of 
$\CM_{\pi_{0}}$. The corresponding 
local $C^*$ algebras, $\CA_{1}(O)$, contain dense subalgebras 
$\CA_{\rm c}(O)$ on which the action of $\alpha_{a}$ is strongly 
continuous in $a$. For every bounded $O$, $\CA_{\rm 
c}(O)^{\prime\prime}=\CA_{1}(O)^{\prime\prime}$ is isomorphic as a 
von Neumann algebra to $\CM_{\pi_{0}}(O)$ because $\pi_{1}$ is 
locally normal. Hence $\{\CA(O)\}$ and 
$\CA_{\rm c}(O)$, which satisfies the hypothesis of 
Theorem 2.1,  have the same family of locally normal 
representations. For the family of these representations one has the
following result:

\begin{theorem} 
Let $\{\CA(O)\}$ be a QFTLO with algebraic PCT symmetry $\theta$. Let $\pi_0$
be a vacuum representation and assume it is invariant under $\theta$.
Then $\theta$ maps the family of locally
normal representations onto itself. Moreover, the family of 
charge sectors is a $\theta$ invariant subfamily of representations.
\end{theorem}

Also the proof of this result will be postponed to the next section.
As a last point we want to look at the algebraic PCT-symmetry in thermal
representations. 
\begin{theorem} 
Let $\{\CA(O)\}$ be QFTLO with PCT-symmetry $\theta$. Assume that the
time translations $\alpha_t$ act strongly continuously. Then the set
of $\beta$-KMS-states is $\theta^*$ invariant.
\end{theorem}

The proof will be given in the next section. 

\subsection{\bf Preparations for the PCT-theorem}
In the QFTLO one usually looks for the PCT-symmetry only in the
vacuum representation of the local observables or of the corresponding 
field algebra.
The requirements for a PCT operator $\Theta$ are:
\begin{itemize}
\item $\Theta$ is antiunitary and for all bounded $O$
\begin{equation}\label{thetacov}\Theta\pi_0(\CA(O))\Theta=\pi_0(\CA(-O)).
\end{equation}
 \item The relation \eqref{thetapoinc} between $\Theta$ and the 
 representation of $\CP_{+}^\uparrow$ holds.
 \item  For field algebras one replaces
$\pi_0(\CA(O))$ by $\CF(O)$ in \eqref{thetacov}. Moreover, it is 
required that $\Theta$ transforms a charge sector into its conjugate 
sector.
\end{itemize}
For the discussion of the PCT-theorem we will mostly look at vacuum
representations. This is sufficient because of Thm.\ 2.2 (see also 
\cite{GL92}). For the 
construction of the PCT-operator on the field algebra one must go
first to the maximal local algebras  and construct the charged fields
as described by Doplicher, Haag and Roberts \cite{DHR69a,DHR69b}, 
\cite{DR90}.
Starting from the PCT-theorem in the vacuum sector one can construct
the PCT-operator for the whole field algebra. This has been done
by Guido and Longo \cite{GL95}, see also \cite{GL92}.

One more remark is in order. Also for the vacuum representation 
of the observable algebra
we will use the phrases PCT-symmetry and PCT-theorem and not simply 
PT-symmetry and so on. The reason is the following: The axioms
of the QFTLO are so general that one can not exclude that the algebras
$\CA(O)$ contain also charged Bose-fields. If they contain such fields
then the group of gauge transformations contains a continuous representation of the circle 
group. But we do not know any manageable condition excluding the existence
of such gauge transformations. If such a charged Bose field is present then one 
has a proper PCT-symmetry also in the vacuum sector.

For a long time it was impossible to prove the PCT-theorem in QFTLO
because of the lack of proper mathematics. This changed with
the appearance of the Tomita-Takesaki theory. The basic result, where
the modular group and conjugation can be computed for some
algebra of interest in QFTLO, was given in \cite{BW75}.

\begin{theorem}{\em (Bisognano and Wichmann).\,}\label{2.4}
Assume a Wightman field theory of a scalar neutral field
is such that the smeared field operators
generate the local algebras. Then:\sabsatz
\begin{itemize}
\item[1.] The modular group of the algebra associated with a wedge and the 
vacuum vector coincides
with the unitary representation of the group of Lorentz boosts which maps the wedge onto itself.
\item[2.] The modular
conjugation of the wedge $W$ is given by the formula
\begin{equation}\label{thetadef}
J_W=\Theta U(R_W(\pi)).
\end{equation}
Here $\Theta$ denotes the PCT-operator of the Wightman field theory
and $U(R_W(\pi))$ is the unitary representation of the rotation which
leaves the characteristic two-plane of the wedge invariant. The angle
of rotation is $\pi$. 
\item[3.] The theory fulfills wedge duality.\end{itemize}
\end{theorem}

In case of the right wedge, $W_{\rm r}=\{x:|x^0|<x^1\}$, the 
group of Lorentz boosts that leave the wedge invariant is 

\begin{equation}\Lambda_{W_{\rm r}}(t)=\left(\begin{array}{cccc}
 \cosh 2 \pi t & -\sinh 2 \pi t &0&0\\
 -\sinh 2 \pi t & \cosh 2 \pi t &0&0\\
 0&0&1&0\\   0&0&0&1
\end{array} \right).\end{equation}
\newline
In particular, $\Lambda_{W_{\rm r}}(-\mfrac\I2)=\newline
{\rm diag\,}(-1,-1,1,1,1)$. 
Moreover, $R_{W_{\rm r}}(\pi)={\rm diag\,}(1,1,-1,-1)$, so  
$\Lambda_{W_{\rm r}}(-\mfrac\I2)R_{W_{\rm r}}(\pi)=-1$.

In a second paper \cite{BW76} Bisognano and Wichmann extended this result 
to
charged Bose and Fermi-fields transforming covariantly with respect 
to a finite dimensional representation of the proper, orthochronous 
Lorentz group in the index space.
The remarkable feature of this result is the fact that modular
group and the modular conjugation of the wedge regions map local 
algebras onto local
algebras. This property has inspired the following 

\begin{definition} \label{def 2.5}  Let $\CM(O)$ be the local algebras 
of a vacuum representation of a QFTLO. 
\begin{itemize}
\item[1.] We say  that
 the {\it Bisognano-Wichmann property} holds in this representation 
 if for any wedge $W$ the modular
group $\Delta_W^{\I t}$ associated with $\CM(W)$ and the vacuum vector acts 
like the corresponding one-parameter group of 
Lorentz boosts $\Lambda_{W}(t)$ that leave $W$ invariant, i.e.
\begin{equation}\label{BW}\Delta_W^{\I t}\CM(O)\Delta_W^{-\I t}=\CM(\Lambda_W(t)O).
\end{equation}
We remark that this property 
has been called {\it modular covariance} 
in \cite{GL95}.
\item[2.] Suppose the
Poincar\'e symmetry is implemented on the representation space
by a unitary representation 
$U(\Lambda,a)$
of $\CP_{+}^\uparrow$. The $U$ is called the {\it minimal
representation} if for every wedge one has
\begin{equation}\Delta_W^{\I t}=U(\Lambda_W(t)).\end{equation}
\end{itemize}
\end{definition}

Since the Lorentz boosts of different wedges generate $\CP_{+}^\uparrow$, 
the minimal representation is unique, when it exists.
For some time it was not known whether the result of Bisognano and
Wichmann holds only for Wightman field theories or if it holds in more 
general cases. A positive answer for the two-dimensional Minkowski
space was given in \cite{Bch92}.

\begin{theorem}{\em (Borchers).}\label{2.6}
Let $\pi_0$ be a vacuum representation of a QFTLO on the two-dimensional
Minkowski space. If the theory fulfills wedge duality, then:
\begin{itemize}
\item[1.] The modular group $\Delta_W^{\I t}$ and the translations 
$U(a)$ generate a representation of the 
two-dimensional Poincar\'e group.
\item[2.] If the local algebras are the maximal algebras, then the
representation defined by the modular group and the translations 
is the minimal representation and 
$\pi_{0}$ has the
Bisognano-property.
\item[3.]The theory {\rm (with maximal local algebras)} is
PCT-covariant. The PCT-operator coincides with $J_W$.\end{itemize}
If wedge duality does not hold for the QFTLO, it is in two dimensions 
always possible to extend the local 
algebras in such a way that it is fulfilled, but the extension is not 
necessarily unique.
\end{theorem}

The basis of this result is the commutation relation between
the modular group and the translations. For the right wedge $W_{\rm 
r}$ then 
in light cone coordinates $x^{\pm}=\frac12 (x^0\pm x^1)$ the relations read
\begin{equation}\label{hstr}
\Delta_{W_{\rm r}}^{\I t}U(x^\pm)\Delta_{W_{\rm r}}^{-\I t}=U(\E^{\mp 2\pi t}x^\pm).
\end{equation}
The spectrum condition for the translations is an essential ingredient
for proving this result, see also \cite{Florig}. Eq.~ \eqref{hstr} holds also 
in higher 
dimension if
$x^\pm$ are the light cone coordinates defined by the two light 
rays $\ell_{1}$, $\ell_{2}$ 
defining the wedge $W$. Also Thm.\ 2.6 can be transcribed to higher
dimensions: Let  $W$ be a wedge satisfying duality, i.e.,  
$\CM(W)'=\CM(W')$. The same holds then for all translates of $W$. 
If $D$ is a double cones in the characteristic
two plane of the wedge and $K(D)$ denotes the cylinder, obtained by translating
$D$ in all directions perpendicular to the characteristic two plane of the 
wedge, then $\Delta_W^{\I t}\CM(K(D))\Delta_W^{-\I t}=\CM(\Lambda_W(t)K(D))$.
But the theorem does not imply that double cones are mapped onto 
double cones. In fact, as known from
the examples of Yngvason \cite{Yng94} (see also \cite{GY00}) the action of the modular group 
of a wedge
does not need to be local in the perpendicular direction.

This result, as well as that of Bisognano and Wichmann, is an 
indication
that wedge duality is important for the proof of the PCT-theorem. In 
fact, it is explicitly or implicitly contained in the hypothesis of
all known proofs, although it is neither a necessary nor sufficient 
condition for PCT symmetry as we will see in the examples in the last section.
In \cite{Bch95} Borchers derived necessary and sufficient conditions for 
the validity of 
wedge duality in QFLTO's with implemented Poincar\'e symmetry. Also here the 
cylindrical sets $K(D)$ just mentioned play an important role, because 
the wedge duality is essentially a two-dimensional problem. 
Before formulating this result let us first discuss the situation. 
This will help to understand the necessary concepts.

Let $W$ be the right wedge
and $\CM(W)$ the corresponding von Neumann algebra 
in a vacuum representation.  This algebra is invariant under the 
representation of the boosts,
$U(\Lambda_W(t))$, and  since this group and the modular group have the vacuum
vector as invariant vector, $U(\Lambda_W(t))$ and $\Delta_W^{\I t}$
commute. Therefore, the two groups differ by a one parameter group $V_W(t)$.
Moreover, if wedge duality holds for $W$, Thm.\ 2.6 
implies that $V_W(t)$ is 
a gauge which maps every $\CM(K(D))$ onto itself. If
$A\in\CM(K(D))$ with $D\subset W$ then $\Delta_W^{\I t}A\Omega$ has an
analytic continuation into the strip $S(-\frac12,0)=\{z\in\BC:-\frac12
<\Imt z<0\}$. At the lower boundary the Tomita-Takesaki theory
implies $\Delta_W^{1/2}A\Omega=J_WA^*J_W\Omega$. If wedge duality
does not hold then the operator
$J_WA^*J_W$ belongs to $\widetilde\CM(K(-D))$ where $\widetilde\CM(K(-D))$
is defined as follows: In the two-dimensional Minkowski space
a double cone is defined by the intersection of two wedges. Therefore,
we can write $K(D)= W_a\cap W'_b$ with $b-a\in W$. So one has
$\CM(K(D))=\CM(W_a)\cap\CM(W'_b)$. With this notation one defines
$\widetilde\CM(K(D))=\CM(W_a)\cap\CM'(W_b)$ which is an algebra 
containing $\CM(K(D))$ as proper subset in the non-duality case.
In this situation $V_W(t)$ maps $\widetilde\CM(K(D))$ onto itself.
If $A\in \CM(K(D))$ with $K(D)\subset W$ it can happen that 
$U(\Lambda_W(t))A\Omega$ also has a bounded analytic continuation into
the strip $S(-\frac12,0)$. If this is the case it can be shown
that there exists an element $\widehat A\eta \CM(K(-D))$ with
\begin{equation}\label{lambdai}
U(\Lambda_W(-\mfrac\I2))A\Omega=\widehat A\Omega.
\end{equation}
The symbol $\eta$ means that the operator might be unbounded but is
affiliated with $\CM(K(-D))$. The group $V_{W}(t)$ is a continuous
Abelian gauge-group. Therefore, if wedge duality holds
the set of elements in $\CM(K(D))$
such that $V_{W}(t)AV_{W}(-t)$ are entire analytic in $t$ is 
{*-strongly} dense
in $\CM(K(D))$. Hence, in this situation the set fulfilling
Eq. \eqref{lambdai} is *-strongly  dense in $\CM(K(D))$. The converse
is the content of the following result.

\begin{theorem}{\em (Borchers).\ }\label{ 2.7}
Assume we are dealing with a vacuum representation of a QFTLO where 
the action of $\CP^{\uparrow}_{+}$ is unitarily implemented.  Assume 
the local algebras are maximal.  The theory fulfills duality for the 
wedges $W,W'$, i.e., $\CM(W)'=\CM(W')$, if and only if for all $D$ 
with $K(D)\subset W$ the set of 
elements $A\in \CM(K(D))$ for which $U(\Lambda_W(t))A\Omega$ has a bounded 
analytic continuation into the strip $S(-\frac12,0)$ is *-strongly 
dense in $\CM(K(D))$.
\end{theorem}

By the results of Bisognano and Wichmann the 
group of the Lorentz boosts and the modular group coincide for the algebra 
of a wedge if the 
QFTLO is generated by Wightman fields transforming covariantly with 
respect to a finite dimensional representation of the Lorentz group.  
In the last theorem we saw that this is a general feature of theories 
in the two-dimensional situation, irrespective whether they are 
generated by Wighman fields or not.  In higher dimensions the modular groups
need not in 
general act like Lorentz boosts, but the assumption that they do
was the starting point of the investigation of Brunetti, Guido and 
Longo \cite{BGL94}.  They obtained the following result:

\begin{theorem}{\em (Brunetti, Guido and Longo).} 
        Let $\{\CM(O)\}$ be the local von Neumann algebras of some 
        representation of a QFTLO and suppose there exists a vector $\Omega$ 
        that is cyclic and separating for all $\CM(O)$ ($O$ open and bounded).
If the Bisognano-Wichmann 
        property \eqref{BW} holds with 
respect to the modular groups defined by the wedge algebras and 
$\Omega$ then the local algebras transform
 covariantly under a representation of the covering group
$\tilde{\CP}^\uparrow_{+}$ of ${\CP}^\uparrow_{+}$. 
This representation fulfills the spectrum 
condition. Moreover, wedge duality holds.
\end{theorem}

The vector $\Omega$ is invariant under the representation of 
$\tilde{\CP}^\uparrow_{+}$ and hence a vacuum vector.  Hence one expects that 
if one deals only with one sector (the vacuum sector) the covering 
group
$\tilde{\CP}^\uparrow_{+}$ can be replaced by ${\CP}^\uparrow_{+}$.  
The solution of this problem is due 
to Guido and Longo \cite{GL95}.  They handled the problem by looking 
at the field algebra $\{\CF(O)\}$ and treating at the same time the 
spin and statistics problem.

\begin{theorem}{\em (Guido and Longo).} 
Assume we are dealing with a representation of a field algebra
$\{\CF(O)\}$. Let there exist 
a vector $\Omega$ in the representation space
which is cyclic an separating for every  $\CF(O)$.
If this representation fulfills the Bisognano-Wichmann property then:
\begin{itemize}
\item[1.] The theory is covariant under a representation of the covering 
$\tilde{\CP}^\uparrow_{+}$ of ${\CP}^\uparrow_{+}$. This representation 
fulfills the spectrum condition.
\item[2.] In the vacuum sector the representation of the Poincar\'e group 
is single-valued and hence it is the minimal representation.
(See Def. 2.5.)
\item[3.]  The standard spin and statistics relations hold.
\item[4.]  The exists a PCT-operator, which transforms the local 
field algebras
correctly, has the correct commutation relations with the Poincar\'e
transformations and maps every sector onto its conjugate sector.
\end{itemize}
\end{theorem}

The question whether one deals in the vacuum sector with a representation
of the Poincar\'e group or its
covering inspired Borchers \cite{Bch96b} to derive 
the minimal group representation directly from the
Bisognano-Wichmann property. In this constructive proof it
is assumed that the local algebras $\CM(O)$ are the maximal algebras.
From the previous discussions it is important to stress the following 
results:
\begin{itemize}
 \item
 {\it The Bisognano-Wichmann property is equivalent to the existence
 of the minimal representation.  These conditions imply wedge
 duality and the PCT-theorem.} 
\end{itemize}

Because of the importance of the modular
groups of the wedges one would therefore like to know whether
other localization properties of these groups are possible, besides 
those described by
Bisognano and Wichmann. This problem has been investigated by 
Kuckert \cite{Ku97}.

\begin{theorem}{\em (Kuckert).} 
Consider a QFTLO which is covariant under translations 
in a vacuum representation. Let $W$ be a wedge,
$D\subset W $  a double cone and $\CM(D)$  maximal.\sabsatz
$1.$ If we assume 
\[
\Delta_W^{\I t}\CM(D)\Delta_W^{-\I t}=\CM(G_{W,D,t})
\]
with some domain $G_{W,D,t}$ then necessarily  $G_{W,D,t}=\Lambda_W(t)D$.\newline
$2.$ If we have
\[
J_W\CM(D)J_W=\CM(\tilde G_{W,D}))
\]
with another domain $\tilde G_{W,D}$, 
then one finds $\tilde G_{W,D}=j_WD$. Here $j_W$ denotes the reflection in the
characteristic two-plane of the wedge which leaves the perpendicular
directions unchanged.
\end{theorem}

It is important to note that these conclusions do not hold in general if the 
equality of the algebras is replaced by an inclusion.
Kuckert assumed that the conditions of the theorem hold for all double 
cones. But this is not necessary because of the translation covariance
and Eq. \eqref{hstr}. This was first observed by Guido \cite{Gui95}.
Recently Kuckert \cite{Ku00} has generalized his theorem by looking at 
individual
operators instead of algebras. Making the additional
assumption of wedge duality and introducing a slightly different notation
of localization he derived the Bisognano-Wichmann
property from the assumption that for every localized operator $A$
the expression $\Delta_W^{\I t}A\Delta_W^{-\I t}$ is localized and this
localization is continuous in $t$. For details see his paper.

\subsection{\bf Implemented Poincar\'e covariance and the PCT-theorem}

Assume one deals with a vacuum representation of a QFTLO which is
covariant under the Poincar\'e symmetry. In this situation it can happen
that the symmetry is implemented by more than one continuous unitary
representation of the Poincar\'e group. If there are two different
representations $U_1(g),U_2(g),g\in \CP^\uparrow_+$ then these
differ by a cocycle $U_2(g)=V(g)U_1(g)$ with values in the gauge group.
$V(g)$ fulfills the cocycle relation
\begin{equation}
V(g_1g_2)=V(g_1)U_1(g_1)V(g_2)U_1(g_1^{-1}).
\end{equation}
Moreover, the unitary family $V(g)$ does not depend on the translations.
If $U_1(g)$ is the minimal representation then $U_1(g)$ and $V(g)$ commute   
for arbitrary $g_1,g_2$. This implies that $V(g)$ itself is a continuous
unitary representation of the Lorentz group \cite{Bch98}. In good situations one
 has only one representation. Such a situation has been described in
\cite{BGL93}.

\begin{theorem}{\em (Brunetti, Guido and Longo).} 
Assume we are dealing with a vacuum representation of an implemented 
Poincar\'e covariant QFTLO. If this representation fulfills the
(distal) split property, then the representation of the Poincar\'e
group is unique.
\end{theorem}

The word ``distal'' means that the  split property explained earlier 
need only be assumed for localization regions sufficiently far from 
each other.  

The last conclusion relies on a result in \cite{DL84} and it
does not imply the existence
of the minimal representation. Therefore, a large part of the investigations
on the PCT-theorem consists in finding conditions for the existence of the 
minimal representation. One such condition can be found in \cite{Bch98}.
Recall from the discussion of the wedge duality that there must be
sufficiently many elements $A\in \CM(K(D))$ with $K(D)\subset W$
such that $U(\Lambda_W(t))A\Omega$ has a bounded analytic continuation into
the strip $S(-\frac12,0)$. For such elements one has
\[
U(\Lambda_W(-\mfrac\I2))A\Omega=\widehat A\Omega
\]
with $\widehat A\eta\CM(K(-D))$. We say $A$ fulfills the {\it reality
condition} if $U(\Lambda_W(t))A^*\Omega$ can also be bounded
analytically continued and if
\begin{equation}
\widehat{A^*}={\widehat A}^*\label{reel}\end{equation}
holds.

\begin{theorem}{\em (Borchers).} \label{2.12}
For a representation of a Poincar\'e covariant theory of local observables
in the vacuum sector the modular group associated with the algebra
of any wedge coincides with the corresponding Lorentz boosts
iff the theory fulfills wedge duality and the set of elements  
$A\in \CM(K(D))$ which fulfill the reality condition Eq. \eqref{reel} is
*-strong dense in $\CM(K(D))$. This implies in particular the 
existence of a PCT operator.
\end{theorem}
        
In a recent paper Guido and Longo \cite{GL00} gave conditions for 
the Bisognano-Wichmann property which have some similarity with
Thm.\ 2.12. Some of the results are more general (statement (iii) and
(iv) of the following theorem)
because the spectrum condition is not required.
\begin{theorem}{\em (Guido and Longo).}\label{2.13}
Let a represented theory of local observables be covariant under a 
representation of the Poincar\'e group $\CP_+^\uparrow$ with an
invariant vector $\Omega$. Let $W$ be a wedge and $S\subset W$ be a
spacelike cone. Assume $\Omega$ is cyclic for $\CM(S)$ and $\CM(W)$. By the 
covariance it is also separating for both algebras.
Let $\CA$ be a weakly dense
*-subalgebra of $\CM(S)$. The following are equivalent:
\begin{itemize}
\item[(i)] The Bisognano-Wichmann relation $\Delta_W^{\frac12}=
U(\Lambda_W(-\mfrac\I2))$ holds.
\item[(ii)] $U(\Lambda_W(-\mfrac\I2)) \CA_1\Omega$ is bounded and the 
translations fulfill the spectrum condition.
\item[(iii)] $U(\Lambda_W(-\mfrac\I2)) \CM(W)_1\Omega$ is bounded.
\item[(iv)] $\norm{U(\Lambda_W(-\mfrac\I2)) \CM(W)_1\Omega}\leq 1$.
\end{itemize}
(The index 1 at the algebras denotes their unit ball.)
Moreover, if the boundary of $S$ intersects the edge $W$ in a half-line, 
then the spectrum condition in (ii) follows already from the boundedness
of $U(\Lambda_W(-\mfrac\I2)) \CA_1\Omega$.
\end{theorem}

We shall now discuss another aspect of the analyticity requirement 
that will lead to a  more direct proof of the PCT theorem in 
Theorem 2.15.
Assume we are dealing with a vacuum representation of a Poincar\'e
covariant theory which fulfills wedge duality and the Bisognano-Wichmann
property. If $D$ is a double cone in a wedge $W$ and away from the 
boundary, then $D$ belongs to many wedges. Let $\Gamma(D)$ be the set of 
Lorentz transformations such that $D\subset gW$. If $A\in\CM(D)$ then
$\Delta_{gW}^{\I t}A\Omega$ has an analytic continuation into
$S(-\frac12,0)$ for every $g\in\Gamma(D)$. Recall that for the wedge
duality it was necessary that there are enough elements in $\CM(K(D))$
such that $U(\Lambda_W(t)$ can be bounded analytically continued 
into $S(-\frac12,0)$. Now we define:
\mabsatz

\begin{definition}\label{2.14}
Let $D$ be a double cone with center at the origin. An element
$A\in\CM(D)$ is called {\it fully analytic} if for any $x$ and every wedge $W$ 
such that $D+x \subset W$ the element $T(x)AT(-x)$ has the above analyticity 
property, i.e., the expression
$U(\Lambda_W(t))T(x)A\Omega$ has a bounded analytic continuation
into $S(-\frac12,0)$.
The set of fully analytic elements of $\CM(D)$ will be
denoted by $\CM^{\rm fa}(D)$.
\end{definition} 

$T(x)$ denotes the representation of the translation.
With this concept we will show

\begin{theorem}\label{2.15}
Let a theory of local observables be covariant under a representation
of the Poincar\'e group which fulfills the spectrum condition.
Assume the fully analytic elements have the following 
properties:
\begin{itemize}
\item[(i)] The set of fully analytic elements is covariant under the
adjoint operation of the Lorentz group, i.e., $A\in \CM^{\rm fa}(D)$ 
implies $U(\Lambda)AU(\Lambda)^{-1}\in \CM^{\rm fa}(\Lambda D)$
\item[(ii)] Let $\CM^{\rm fa}(W)=\cup\{T(x)\CM^{\rm fa}(D)T(-x):D+x\subset W\}$
then $\CM^{\rm fa}(W)\Omega$ is a core for $U(\Lambda_W(-\frac\I2))$
as well as for $\Delta_W^{1/2}$.
\end{itemize}
If these conditions are fulfilled then a PCT-operator exists.
\end{theorem}

We remark that by Nelson's theorem 
$\CM^{\rm fa}(W)\Omega$ is a core  for $U(\Lambda_W(-\frac\I2))$ if this 
domain is dense in the Hilbert space and invariant under the Lorentz 
boosts. It is a core for $\Delta_W^{1/2}$ if $\CM^{\rm fa}(W)$ is 
*-strongly dense in $\CM(W)$.

The proof of this result will also be postponed to the next section.
One other result concerning the fully analytic elements is the following:

\begin{theorem} \label{ 2.16}
Let a theory of local observables be covariant under a representation
of the Poincar\'e group fulfilling the spectrum condition.
Then the assumptions of the last theorem hold if and only if the
given representation of the Poincar\'e group is the minimal one. This 
implies that every localized element is fully analytic.
\end{theorem}

Also the proof of this result will be given in the next section.

One obtains a direct application of the result of 
Bisognano and Wichmann \cite{BW76} if one deals with a QFTLO
with isolated masses giving rise to complete asymptotic fields. In this
situation J. Mund \cite{Mu00} has shown that the modular groups of any wedge
coincide for the interacting and the asymptotic fields. This result
holds also if the charged fields are localized in space like cones as
described by Buchholz and Fredenhagen \cite{BF82}.
The proof is possible since one knows the the commutation relations
between the modular transformations and the translations \eqref{hstr}.

\begin{theorem}{\em (Mund).}\label{2.17}
Assume we are dealing with a representation of a field algebra
$\{\CF(O)\}$ which is covariant under a representation of the connected
Poincar\'e group. Let there exist a vacuum vector $\Omega$ in the 
representation space which is cyclic an separating for every  $\CF(O)$ 
and let the theory fulfil (twisted) wedge duality. Assume moreover, 
that the spectrum of the translations has isolated masses
in the sectors describing elementary charges, 
and that the corresponding asymptotic fields have finite multiplicity and 
are complete. Then the theory has the Bisognano-Wichmann property.
\end{theorem}

\subsection{\bf Supplements to the PCT-theorem}

In the previous investigations we looked at properties of the modular
groups of the wedge algebras. Therefore, some results can be obtained
by looking at wedges alone. Buchholz and Summers initiated a program
where they only used the modular conjugations of all wedges \cite{BS93}.
The idea is based on the well known fact that the Poincar\'e group $\CP_+$
is generated by reflections. Therefore, the modular conjugations $J_W$
should generate a representation of $\CP_+$. The original Ansatz has been
generalized considerably in recent papers \cite{BDFS00} and 
\cite{BFS99}.
In order to discuss their results we must introduce the concepts used.
(We restrict ourselves to the case of the four-dimensional Minkowski space.)

By $\CW$ we denote the set of all wedges (including the translated ones).
For $W\in\CW$ there shall exist a bijection $\tau_W:\CW\to\CW$ with
\begin{itemize}
\item[(i)] $\tau_W^2=id$.
\item[(ii)] $W_1\subset W_2\to \tau_W(W_1)\subset\tau_W(W_2)$.
\item[(iii)]$\overline W_1\cap\overline W_2=\emptyset\to
  \tau_W(W_1)\cap
\tau_W(W_2)=\emptyset$.
\end{itemize}
If these conditions are fulfilled then one concludes \cite{BDFS00}:
\begin{itemize}
\item[1.] $\tau_W(W'_1)=(\tau_W(W_1))'$.
\item[2.]To $\tau_W$ there exists an element $g$ of the Poincar\'e group with
$\tau_W(W_1)=\alpha_g(W_1)$.
\item[3.] If the group generated by the $\tau_W$ acts transitively on $\CW$
and if for one $W$ there holds $\tau_W W=W'$ then the group generated
by the $\tau_W$ is $\CP_+$.
\end{itemize}
Next we look at the theory defined by the wedges. Let $\{\CM(W).\Omega\}$
be a representation such that $\Omega$ is cyclic for every $\CM(W)$.
Assume to every $W$ exists an anti-unitary operator $J_W$ such that
\[
J_W\CM(W_1)J_W=\CM(\tau_W(W_1))
\]
holds. Moreover it is required that
\begin{itemize}
\item[(iv)] $W\to\CM(W)$ is an order preserving bijection.
\item[(v)] $W_1\cap W_2\neq\emptyset\to \Omega$ is cyclic for 
$\CM(W_1)\cap\CM(W_2)$. If $\Omega$ is cyclic for $\CM(W_1)\cap\CM(W_2)$
then holds $\overline W_1\cap\overline W_2\neq\emptyset$.
\end{itemize}
Using this the authors \cite{BDFS00} obtained:

\begin{theorem}{\em (Buchholz, Dreyer, Florig and Summers).}\label{2.18}
Assume we are dealing with a family of wedge algebras on a
Hilbert space $\CH$ and mappings $\tau_{W}$, such that the conditions (i)--(v) above are 
fulfilled. Let a vector $\Omega\in\CH$ be cyclic for every
$\CM(W)$. Assume to every $W$ there exists an antilinear involution $J_W$ such 
that
\[
J_W\CM(W_1)J_W=\CM(\tau_WW_1).
\]
Furthermore, assume $J_W\Omega=\Omega$
for every $W$ and that $\{\tau_W\}$ has the transitivity property mentioned 
under 3 above. Then the operators $J_W$ generate a projective representation
of the group $\CP_+$. If we denote by $J(g)$ the representant of $g\in\CP_+$
then
\[
J(g)\CM(W)J^*(g)=\CM(gW).
\]
Moreover, the theory fulfills wedge duality, and since $\CP_+$ contains the
element $-1$ the theory is PCT-covariant with $\Theta=J(-1)$.
\end{theorem}

Since by assumption (ii) the $\CM(W)$ generate the algebras of the
double cones one obtains Poincar\'e covariance of the QFTLO. In
\cite{BFS99} it was shown that the representation of the translation
group is continuous. But the representation of the translations does
not need to fulfill the  spectrum condition. The spectrum condition holds only
if the group generated by the $J_W$ contains also the modular
groups of the wedges.

Another aspect of the wedge algebras and the modular theory is the
construction of a QFTLO out of a finite set of wedge algebras
fulfilling some requirements among each other. This construction leads
to the family of all wedge algebras. This set is covariant
under a representation of the Poincar\'e group $\CP^\uparrow_+$. Since 
this representation is generated by the modular groups of the wedges
it is the minimal representation. These results are due to Wiesbrock
\cite{Wie97,Wie98} and to K\"ahler and Wiesbrock \cite{KW99}. 
We do not want to go into 
the details of this program because this would need the introduction
of too many new concepts.

\section{Proofs}

In this section we present the proofs of the new results, Thms.\ 
2.1--2.3 and Thms.\ 2.15 and \ 2.16. 
For the proofs of Thms.\ 
2.1--2.3 one should notice that the antilinearity of $\theta$
implies the relation $(\theta^*\omega)(A)=\omega(\theta(A^*))$. This is
best seen by looking at a
situation where $\theta$ is implemented. In this case one has
\[
\omega_\psi(\theta(A))=(\psi,\Theta A\Theta \psi)=
(\Theta \psi,A^*\Theta\psi)=(\theta^*\omega_{\psi})(A^*).
\]

\begin{proof}{\bf of Theorem 2.1}
For $f\in \CL^1(\BR^d)$ and $A\in\CA(\BR^d)$
set $A(f)=\newline \int d^d xf(x)\alpha_x(A)$. This is defined as a strong
integral since $\alpha_x$ acts strongly continuously. For $a\in V^+$
let $\CL_a$ be the left-ideal defined by the set
\[
\{A(f):A\in\CA(\BR^d),f(x)\in\CL^1(\BR^d),\;{\rm and\ support}\;
\tilde f(p) \subset \BR^d\setminus (-a+\overline V^+)\}.
\] 
If $\omega$ is a state on $\CA(\BR^d)$ with $\omega(\CL_a)=0$
then $\pi_\omega$ defines a particle representation. If $E(p)$ denotes
the spectral family of the corresponding representation of the translation 
group one has $\omega(E(D_{0,a})=1$ where $D_{0,a}=V^+\cap( -V^+a)$. Conversely we know
that the set of all states, fulfilling $\omega(\CL_a)=0$ for some 
$a\in V^+$, is norm-dense in the set of all normal states (folium)
of the universal particle representation.

Since $\theta$ and the translations fulfill \eqref{thetaalpha}, we 
have
\[
\begin{split}
\theta(A(f))&=\theta(\int dx\{\int dp\, \E^{\I px} \alpha_x(A)\})\\
&=\int dx\{\int dp\, \E^{-\I px} \alpha_{-x}(\theta(A))\}=(\theta(A))(f).
\end{split}
\]
Hence we get $\theta(\CL_a)=\CL_a$. This implies $\theta^*$ maps a 
norm-dense set
of normal states of the universal particle representation onto itself.
This implies the invariance of the set of particle representations.

The vacuum states are characterized by the annihilation of $\CL_0$. This 
ideal is $\theta$ invariant, implying the invariance of the vacuum
representations. 

For details on the universal particle representation see \cite{Bch96}. The
left-ideals $\CL_a$ were introduced in \cite{Bch70}.
\end{proof}

\begin{proof}{\bf of Theorem 2.2}
Let $\omega$ be a locally normal state on
$\CA(\BR^d)$ with respect to a vacuum representation $\pi_0$. Since this 
representation is $\theta$-invariant there exists an anti-unitary operator
$\Theta$ on $\CH_0$ with 
\[
\Theta\pi_0(A)\Theta=\pi_0(\theta A).
\] 
Since $\pi_0$ is a vacuum representation it has the Reeh-Schlieder
property. Consequently every normal state on $\pi_0(\CA(O))$ is a 
vector state. In particular local normality implies that for
every $O$ there exists a vector $\psi(O)$ with
$\omega(A)=(\psi(O),\pi_0(A)\psi(O))$ for all $A\in\CA(O)$. 
Let now $O$ be symmetric
with respect to the origin. Then $A\in\CA(O)$ implies
$\theta(A)\in\CA(O)$ and then obtain
\[
\begin{split}
(\theta^*\omega)(A)=\omega(\theta(A^*))&=(\psi(O),\pi_0(\theta(A^*))\psi(O))
=\\(\psi(O),\Theta\pi_0(A^*)\Theta\psi(O))&=
(\Theta\psi(O),\pi_0(A)\Theta\psi(O)).
\end{split}
\]
Hence $(\theta^*\omega)$
is a normal functional on $\pi_0(\CA(O))$. This implies the first statement.

For the second statement notice that in a particle representation
the translations are unitarily implemented
and we know the commutation relations \eqref{thetaalpha} of $\theta$ with the translations.
Since every double cone can be translated in such a way that its
center is at the origin we find that $(\theta^*\omega)$ is normal
on $\pi_0(\CA(O'+x))$ if and only if it is normal on $\pi_0(\CA(O'))$.
Hence it is sufficient to look at the complement of symmetric
double cones. Since the algebras if these sets are invariant under
$\theta$ the method of the locally normal case can be applied. Consequently
the set of particle representations is $\theta$-invariant. 
\end{proof}

\begin{proof}{\bf of Theorem 2.3}
Let $\omega$ be a $\beta$-KMS-state,
then for $A,B\in\CA(\BR^d)$ the expression $F(t)=\omega(B\alpha_t(A))$
can be analytically continued into $S(-\beta,0)$. At the lower boundary
this function has the value $F(t-i\beta)=\omega(\alpha_t(A)B)$. 
Since the algebra is translation invariant and since the translations
act strongly continuously, there is a norm-dense sub-algebra
$\CA^{{\rm an}}(\BR^d)$ such that for $A\in\CA^{{\rm an}}(\BR^d)$
the expression $\alpha_t(A)$ is entire analytic in $t$. From the
relation $\alpha_t\circ\theta=\theta\circ\alpha_{-t}$ we see that
$\CA^{{\rm an}}(\BR^d)$ is invariant under $\theta$. Moreover, the 
antilinearity of $\theta$ implies for complex $z$ the equation
\[
\alpha_z(\theta(A))=\theta(\alpha_{-\overline z}(A)).
\]
Inserting for $A,B\in\CA^{{\rm an}}(\BR^d)$ the operators
$\theta(A)$ and $\theta(B)$ into the expression for 
$F(t)$ one obtains for complex $z$
\[
F(z)=\omega(\theta(B)\alpha_z(\theta(A)))=\omega(\theta(B)
\alpha_{-\overline z}(A)))=(\theta^*\omega)(\alpha_{-z}(A^*)B^*).
\]
Since $\theta$ commutes with the time translations it follows that also
$\theta^*\omega$ is translation invariant. Hence we obtain
\[ 
(\theta^*\omega)(\alpha_{-z}(A^*)B^*)=
(\theta^*\omega)(A^*\alpha_{z}(B^*)).
\]
From
\[
\begin{split}
F(t-\I\beta)=\omega(\alpha_t(\theta(A))\theta(B))&=
\omega(\theta(\alpha_{-t}(A)B))=\\
(\theta^*\omega)(B^*\alpha_{-t}(A^*))&
=(\theta^*\omega)(\alpha_t(B^*)A^*),
\end{split}
\]
we see that $(\theta\omega)$ fulfills the $\beta$-KMS-condition for 
elements $A,B\in\CA^{{\rm an}}(\BR^d)$. Since $\CA^{{\rm an}}(\BR^d)$
is norm-dense in $\CA(\BR^d)$, the $\beta$-KMS-conditions holds for
arbitrary elements. This implies that $\theta^*\omega$ is again a 
$\beta$-KMS-state.  Hence
the set of $\beta$-KMS-representations is $\theta$ invariant.
\end{proof}
\medbreak
{\bf Proof of Theorem 2.15}
If the representation of the Poincar\'e
group is the minimal one, then Eq.\ \eqref{thetadef} holds, i.e.,
\[
\Theta=J_W U(R_W(\pi)),
\]
provided the origin is contained in the edge of the wedge. $R_W(\alpha)$
denotes the rotation in the two-plane perpendicular to the
characteristic two-plane of the wedge, and $J_W$ the modular
conjugation of the algebra of the wedge.
Therefore, one has to solve two problems:

We know from Thm.\ 2.7 that the wedge duality is equivalent to analyticity 
properties of sufficiently many $A\in\CM(D),\; D\subset W$ and this 
condition is in particular fulfilled by our assumption (ii). For these 
elements
\[ 
U(\Lambda_W(-\mfrac\I2))A\Omega=\widehat A\Omega ,
\]
holds with $\widehat A\eta \CM(K(P_W D))$,
where $K$ is a cylindrical set introduced in the remarks following 
Thm.\ 2.6 and $P_W$ the reflection in the characteristic two-plane of the
wedge.
In order to show the PCT-theorem one must prove that the map
\[
A\Omega\longrightarrow \widehat A\Omega,\quad A\in\CM(D),\; D\subset W
\]
sends double cone algebras into double cone algebras and not only into 
cylindrical set algebras.

The problem of the double cones can be looked at as follows:
The element $R_W(\pi)\Lambda_W(-\frac{\I}2)$ considered as an element
of the complex Lorentz group is the unique element $-1$. Looking
at representations, then $U(\Lambda_W(t)$ has an analytic continuation
when applied to vectors $A\Omega$ with sufficient analyticity properties.
Starting from different wedges it must be shown that one does not end
by $-1$ on different sheets of an analytic manifold.

The second problem is the following: From the discussion following Thm.\ 2.6
we know that one can write
\[
U(\Lambda_W(t))=\Delta_W^{\I t} V_{W}(t),
\]
where $V_{W}(t)$ maps every translated cylinder set algebra 
$\CM(K(D)+x)$, $D\in W$ onto itself. We would like to know 
conditions implying that $V_{W}(t)$ is a gauge transformation, which means that
it maps the algebra of every double cone onto itself.

We start with the first problem. 
Let $D$ be a double cone with center at the origin. Let
$D+x\subset W$ and  define
\begin{equation}\label{gamma}
\Gamma(D+x)=\{g\in \CP^{\uparrow}_+:\; \overline D+x\subset gW\}.
\end{equation}
Since $W$ is open, $\Gamma(D+x)$ is open and contains the identity of the
group.
The first observation is:

\begin{lemma}
There exist $g_1,g_2,...,g_6\in \Gamma(D+x)$ and $T_1,...T_6>0$, such that
\[
D+x\subset \Lambda_{g_6 W}(t_6)\cdots\Lambda_{g_1 W}(t_1) W
\]
for $|t_i|<T_i,\; i=1,...,6$. The elements $g_1,...,g_6$ can be chosen
in such a way that the generators of the groups $\Lambda_{g_i W}(t_i)$
are linearly independent.
\end{lemma}

\begin{proof}
Let a neighborhood of the identity
$\Gamma_1(D+x)$ be a subset of $\Gamma(D+x)$ such that 
$g_1,...,g_6\in \Gamma_1(D+x)$
implies $g_1...g_6\in \Gamma(D+x)$. Since $\Gamma_1(D+x)$ is a neighborhood of
the identity there exists $g_1=1,g_2,...,g_6\in \Gamma_1(D+x)$, such that the
generators of $\Lambda_{g_iW}(t)$ are linearly independent. Choosing
$T_i$ such that $\Lambda_{g_iW}(t_i)\in \Gamma_1(D+x)$ for $|t_i|<T_i$
then the statements of the lemma are fulfilled. 
\end{proof}

With help of the last lemma we can construct an analytic function
on parts of the complex Lorentz group $\hat \CP_+$. (Its elements
will be denoted by $\hat g$.)

\begin{proposition}
Let $A\in \CM^{\rm fa}(D)$ and let $g_1,\dots,g_6$ and 
$D+x$ be as in the last lemma. Then the function
\begin{equation}\label{prod}
U(\Lambda_{g_6W}(t_6))...U(\Lambda_{g_{1}W}(t_1))A\Omega,\quad A\in
\CM(D+x),\quad D+x\subset W
\end{equation}
has an analytic continuation into all $t$--variables. The function
$U(\Lambda_{g_jW}(-\frac\I2))A\Omega$ 
is the boundary value of an analytic function and the product
\begin{equation}\label{u-1}
U(R_{g_jW}(\pi))U(\Lambda_{g_jW}(-\mfrac\I2))A\Omega
\end{equation}
is independent of $g_j$. 
\end{proposition}

\begin{proof}
In the variable $t_i$ the above function can be 
analytically extended in the  into the strip $S(-\frac12,0)$.
provided we keep $t_6,...,t_{i+1},t_{i-1},...,t_1$ real and in their
proper domain. An analytic continuation in all $t$ variables is
obtained with help of the Malgrange--Zerner theorem (see e.g. \cite{Ep66})  
The domain into which the function can be continued has
still to be determined. It is clear from the construction
that the real function is the boundary value of the analytic continuation.

Next we want to determine the domain of holomorphy of this function.
This calculation will be done
by mapping the strip $S(-\frac12,0)$ bi--holomorphically onto itself
in such a way that the interval $|x|<T$ is mapped onto $\BR$
and the rest of the boundary onto $-\frac\I2+\BR$. This is achieved
by the transformation
\begin{equation}\label{zeta}
\zeta=\frac{1}{2\pi}\log\frac{1-\E^{-2\pi T}}{\E^{2\pi T}-1}
\frac{\E^{2\pi T}-\E^{-2\pi z}}{\E^{-2\pi z}-\E^{-2\pi T}}.
\end{equation}
With these new variables the domain of holomorphy becomes
\begin{equation}\label{domain}
0>\sum\limits_{i=1}^6 \Imt\zeta_i>-\frac12.
\end{equation}
If the elements $g_1,...,g_6$ are properly chosen
then an interior point of the $\zeta$ variables corresponds to an
interior point in the $\hat g$ variables.

In the $\zeta$--variable the domain \eqref{domain} is 
convex and hence simply
connected. Since the transformation \eqref{zeta} 
is bi--holomorphic, it follows 
that also the image in the $t$--variables is simply connected. Hence 
there are no monodromy problems in these variables. 

Note that the symbol $U(\Lambda)$
denotes a representation of the Lorentz group and therefore, the first
expression of Eq.~ \eqref{prod} can be written as
\begin{equation}\label{prod2}
U(\Lambda_{g_6W}(t_6)\cdots\Lambda_{g_{1}W}(t_1))A\Omega.
\end{equation}
The arguments inside of $U$ are defined for all elements of the
complex Lorentz group. $U$ applied to this product is defined if it belongs
to the domain Eq. \eqref{domain} (transformed into the variable $z$), 
eventually multiplied from the left with an element of the real
Lorentz group.

In particular we can look at the product 
$U(R_{g_jW}(\pi))U(\Lambda_{g_jW}(-\frac\I2))U(g)A\Omega$
for  $g$ sufficiently close to $1$ in \eqref{gamma} and obtain
\[
\begin{split}
U(R_{g_jW}(\pi)\Lambda_{g_jW}(-\mfrac\I2)&g)A\Omega=
U(gR_{g_jW}(\pi)\Lambda_{g_jW}(-\mfrac\I2))A\Omega=\\ &
U(g)U(R_{g_jW}(\pi))U(\Lambda_{g_jW}(-\mfrac\I2))A\Omega.
\end{split}
\]
This implies
$U(g)U(R_{g_jW}(\pi))U(\Lambda_{g_jW}(-\frac\I2))U(g^{-1})
=\newline U(R_{gg_jW}(\pi))U(\Lambda_{gg_jW}(-\frac\I2))$.
Choosing $\Gamma_1(D+x)$ in such a way that it contains with $g_j$ 
also its inverse, then the statement of the proposition is obtained by
choosing $g=g_ig_j^{-1}$.
\end{proof}

Collecting the result of the discussion we obtain with $A_x=T(x)AT(-x)$

\begin{proposition}
Let $D$ be a double cone centered at the origin and 
$x$ such that the closure of $D+x$ does not
contain the origin. Then for $A\in\CM^{\rm fa}(D)$ and $g$ such that
$D+x\subset gW$ the vector function
\[
U(\Lambda_{gW}(-\mfrac\I2))U(R_{gW}(\pi))A_x\Omega
\]
is independent of $g$.
\end{proposition}

\begin{proof}
From the above discussion we know that the statement is
true for $g$ in a sufficiently small neighborhood of the identity
in $\CP^\uparrow_+$. But this implies by varying the wedges and the
neighborhoods that it is true for all $g\in \Gamma(D+x)$.
\end{proof}

Using Prop. 3.3 we find

\begin{theorem}
Let $A\in \CM^{\rm fa}(D)$ with $D$ centered at the origin, and let $D+x\subset W$. Then one has
\[
U(\Lambda_W(-\mfrac\I2))A_x\Omega=\widehat A_{P_W x}\Omega
\]
with $\widehat A_{P_W x}\in \CM(D+P_W x)$.
\end{theorem}

\begin{proof}
From Eq. (3.4) we know $U(R_{gW}(\pi))
U(\Lambda_{gW}(-\mfrac\I2))A_x\Omega$ 
is independent of $g$ as long as $g$ belongs to the set
$\Gamma(D+x)$ (see Eq. (3.2)). Hence we obtain
\[
U(R_{W}(\pi))U(\Lambda_{W}(-\mfrac\I2))A_x\Omega=U(R_{W}(\pi))
\widehat A_x\Omega
\]
with
\[
U(R_{W}(\pi)\widehat A_x U(R_{W}(\pi)\subset\mathop{\cap}
\limits_{g\in \Gamma(D+x)} \CM(-K_{gW}(D+x)).
\]
Next observe that a translation in the characteristic two-plane of $W$ 
commutes with $U(R_{W}(\pi))$ and is mapped onto its negative by
$U(\Lambda_{W}(-\frac\I2))$. This implies
\[
U(R_{W}(\pi))\widehat A_x U(R_{W}(\pi))\subset\mathop{\cap}
\limits_{D+x\subset W}\mathop{\cap}
\limits_{g\in \Gamma(D+x)} \CM(-K_{gW}(D+x)=\CM(-D-x).
\]
Using the fact that $D$ is symmetric and that $R_W(\pi)(-x)=P_Wx$ holds
we get the result of the theorem.
\end{proof}

Now we are prepared to show Thm. 2.15. 
\begin{proof}
First a remark: If $\CA\in\CM(W)$ and $U(\Lambda_W(t)A\Omega$ has an
analytic continuation into the strip $S(-\frac12,0)$ then the same holds
for $V_W(t)A\Omega$ because $\Delta_W^{\I t}A\Omega$ has an analytic
continuation and the relation 
\begin{equation}\label{gauge}
U(\Lambda_{W}(t))=V_W(t)\Delta_W^{\I t}
\end{equation}
holds. (For a detailed proof see \cite{Bch96}.) Since all groups in
\eqref{gauge} leave the wedge $W$ invariant it follows that they commute.
If $A\in\CM(D)^{\rm fa}$
then $U(R_{gW}(\pi))U(\Lambda_{gW}(-\frac\I2))$ acts locally on
$A_x\Omega$
i.e., it maps $A_x\Omega$ into $\CM(D-x)$ 
and is independent of $gW$, as long as $g\in\Gamma(D+x)$. By
Eq.\ \eqref{gauge} we can
replace $U(\Lambda_{gW}(-\frac\I2)$ by $R_{gW}(-\frac\I2)
\Delta_{gW}^{\frac12}$. This implies that $U(R_{gW}(\pi))V_{gW}(-\frac\I2)
\Delta_{gW}^{\frac12}$ acts locally on
$A_x\Omega$ and is independent of $g$, as long as $g\in\Gamma(D+x)$.
Replacing $A_x$ by its adjoint $A_x^*$ and observing that the Tomita 
conjugation $S_{gW}$ acts locally on 
$A_x\Omega$ and is independent of $g$ in the same range as before, we find,
that $U(R_{gW}(\pi))V_{gW}(-\frac\I2)J_{gW}$ has the good properties, 
i.e., it maps $A_x\Omega$ into $\CM(D-x)\Omega$ and is independent of $g$ 
for $g\in\Gamma(D+x))$. We define
\[
\widehat\Theta_{W}=U(R_{W}(\pi))V_{W}(-\mfrac\I2)J_{W}=
U(R_{W}(\pi))J_{W} V_{W}(\mfrac\I2).
\]
Notice that $J_W$ commutes with $V_W(t)$. Hence we obtain 
$J_W V_W(-\frac\I2)J_W=V_W(\frac\I2)\newline
=V_W(-\frac\I2)^{-1}$. This implies
${\widehat\Theta_W}^2={\bf 1}$. 
By assumption the fully analytic elements applied to the vacuum are a
core for $\Delta_W^{-1/2},U(\Lambda_W(-\frac{\I}2)$ and hence also for
$V_W(-\frac\I2)$ and $\widehat\Theta_W$. (Remember that
$U(\Lambda_W(t))V_W(t)=\Delta_W^{\I t}$.)
Denoting the closure by the
same symbol, $\widehat\Theta_W$ has a polar decomposition
\[
\widehat\Theta_W=\Theta_W T_W.
\]
By the uniqueness of the decomposition and the positivity of
$V_W(\frac\I2)$ it follows that the relation $T_W=V_W(\frac\I2)$ holds.
Since $\widehat\Theta_W$
maps $\CM^{\rm fa}(D+x)$ onto $\CM^{\rm fa}(D-x)$ the same must hold for
$\Theta_W$. Since by analytic continuation these expression are
independent of $W$ we get that $\Theta_W=\Theta$ is the PCT-operator
and $V_W(\frac\I2)$
is independent of $W$ and maps $\CM^{\rm fa}(D+x)$ onto itself.
\end{proof}

\begin{proof}{\bf of theorem 2.16}
During the proof of the last theorem we had
constructed the operator
$\widehat\Theta=U(R_W(\pi))J_W V_W(\frac\I2)$. By the uniqueness of
the polar decomposition we see that the positive operator $V_W(\frac\I2)$
and the PCT-operator $\Theta=U(R_W(\pi))J_W$ act locally and are
independent of $W$. This implies that $J_W$ acts locally and maps
$\CM(D+x)$ onto $\CM(D+P_W x)$. Writing $V_W(t)= \E^{\I X_W t}$ then 
since
$\E^{- X_W /2}$ is independent of $W$ we conclude that also
$V_W(t)= \E^{\I X_W t}=V(t)$ is independent of $W$. This implies 
\[
[V(t),U(\Lambda)]=0
\]
for all $\Lambda$.
We know that $V(t)$ maps $\CM(K_W(D+x))$ onto itself for every $W$. Choosing 
$x=0$ and varying $W$ we find that $V(t)$ maps $\CM(D)$ onto itself. Since
$V(t)$ commutes with the translations it is a local gauge. Hence
also $\Delta_W^{\I t}$ acts locally for every $W$. Therefore
$V(t)$ must define a representation of the Lorentz group. Since this group
representation is Abelian it must be trivial. This implies that $U(g)$ is
the minimal representation. 
\end{proof}
           
\section{Examples}
Free quantum fields, i.e., fields that fulfill the Klein Gordon 
equation,  provide simple examples illustrating several of 
the points discussed in the previous sections. 

We consider first the case of a single, hermitian Bose field $\Phi$  
with mass $m>0$
on $\BR^d$, $d\geq 2$, transforming covariantly with 
respect to the translation group, but not necessarily w.r.t.\ the 
Lorentz group.  The Fock space representation of $\Phi$ is determined 
by the two point function,
\begin{equation}
\mathcal W_{2}(x-y)=\langle\Omega,\Phi(x)\Phi(y)\Omega\rangle.
\end{equation}
Locality, spectrum condition and positivity imply (by the
Jost-Lehmann Dyson representation \cite{Bch96}) that its Fourier transform
can be written
\begin{equation}
\tilde {\mathcal W}_{2}(p)=M(p)\theta(p^0)\delta(p\cdot p-m^2)
\end{equation}
where $M(p)$ is a polynomial in $p$ satisfying
\begin{equation}
M(p)=M(-p)\quad\text{and}\quad M(p)\geq 0     
\end{equation}
for $p$ on the positive mass shell $H_{m}^+=\{p:\ p^0>0,\ p\cdot 
p=m^2\}$.
Conversely, every such $M$ defines a free field satisfying all 
Wightman axioms except possibly Lorentz covariance.
The field $\Phi$ gives rise to a QFTLO where the local von Neumann 
algebras $\CM(O)$ are generated by the Weyl operators 
$\exp(\I\Phi(f)$ with real test functions $f$ supported in $O$.

According to the results of \cite{GY00} wedge duality is always 
violated unless $M$ is constant on the mass shell, i.e., unless 
$\Phi$ is the usual Poincar\'e covariant, scalar free field. In 
particular, neither the condition of geometric modular action nor the 
Bisognano Wichmann property hold for fields with non-constant $M$. On 
the other hand, it is trivial that PCT symmetry {\it always} 
holds in these examples, 
irrespective of $M$ and $d$. In fact, since $\Phi(x)=\Phi(x)^*$ and 
$\mathcal W_{2}$ depends only on $x-y$, we have
\begin{equation}
    \langle\Omega,\Phi(x)\Phi(y)\Omega\rangle
    =\langle\Omega,\Phi(-y)\Phi(-x)\Omega\rangle
    \label{freepct}
\end{equation}
 which is exactly PCT symmetry. Space and time inversion  are only 
symmetries separately, however, if $M$ is even in $p^0$. These examples thus shows 
that
\begin{itemize}
 \item Wedge duality is not a necessary condition for PCT symmetry.
 \item Full PCT symmetry is not in conflict with odd dimensionality 
 of space-time.
\end{itemize}

More generally we may consider free fields $\Phi_{\alpha}$ that have
an arbitrary number of components and are not necessary hermitian. As 
before we assume translation covariance and spectrum condition, but 
not necessarily Lorentz covariance.
It is convenient to take the set of indices $\alpha$, that label
the components of the field including the adjoint operators, as a
basis of a complex vector space $\CK$.  The field operators
$\Phi(f,\sigma)$ thus depend linearly on $\sigma\in\CK$ besides the
test function $f$, and we can write the adjoints as
$\Phi(f,\sigma)^*=\Phi(\bar f,\sigma^*)$, where $\bar f$ is the
complex conjugate test function and $\sigma\to\sigma^*$ is an
antilinear involution on $\CK$.  (In the case of hermitian fields
$\sigma^*$ is simply complex conjugation of the components of $\sigma$
with respect to the basis.)  The Fourier transform of the two point
function
\begin{equation}
   \mathcal 
   W_{2}(x-y;\sigma,\rho)=\langle\Omega,\Phi(x,\sigma)\Phi(y,\rho)\Omega\rangle
\end{equation}
can now be written as
\begin{equation}
\tilde {\mathcal W}_{2}(p;\sigma,\rho)=M_{\sigma,\rho}(p)\theta(p^0)\delta(p\cdot p-m^2)
\end{equation}
where $M_{\sigma,\rho}(p)$ is a polynomial in $p$ that depends 
bilinearly on $\sigma$ and $\rho$. For Bose fields locality is 
equilvalent to 
\begin{equation}
    M_{\sigma,\rho}(p)=M_{\rho,\sigma}(-p)
    \label{locality}
\end{equation}
and positivity to
\begin{equation}
    M_{\sigma^*,\sigma}(p)\geq 0,\qquad\text{for}\ p\in H_{m}^+,
    \label{posit}
\end{equation}
which implies in particular the hermiticity condition
\begin{equation}
    M_{\sigma^*,\rho}(p)=\overline{M_{\rho^*,\sigma}(p)}
        ,\qquad\text{for}\ p\in H_{m}^+.
    \label{hermit}
\end{equation}
PCT symmetry of the field, on the other hand, requires that
\begin{equation}
    M_{\sigma,\rho}(p)=\overline{M_{\rho,\sigma}(p)}=M_{\sigma^{*},\rho^{*}}(p).
    \label{pctsymmfield}
\end{equation}
It is easy to give examples that satisfy \eqref{locality} and \eqref{posit}
but not \eqref{pctsymmfield}. For instance, on can take an $n$ 
component 
field, $n\geq 2$, with $M(p)$ given by a positive definite  
$n\times n$ matrix whose diagonal elements are even polynomials in $p$ 
and whose off diagonal elements are odd polynomials that 
are
purely imaginary for 
$p\in H_{m}^+$. It has to be remarked, however, that not
all such examples violate PCT symmetry {\it for the algebra of 
observables} 
generated by the field. In fact, two different multicomponent fields 
$\Phi_{\alpha}^{(1)}$ and  
$\Phi_{\beta}^{(2)}$ may generate the same local algebras of 
observables, and one of them can fulfill \eqref{pctsymmfield} while the 
other 
does not. To discuss this in a little more detail let 
us equip $\CK$ with a scalar product $\langle\cdot,\cdot\rangle$ 
and write 
\begin{equation}
    M_{\sigma^{*},\rho}(p)=\langle\sigma, M(p)\rho\rangle
    \label{moperator}
\end{equation}
where $M(p)$ is a hermitian linear operator on $\CK$. If $M^{(1)}(p)$  
and $M^{(2)}(p)$ correspond to the two different fields, then the 
generated local algebras are clearly equal if
\begin{equation}
    M^{(1)}(p)=L(p)^{*} M^{(2)}(p)L(p)
    \label{equalalgebras}
\end{equation}
where $L(p)$ is  an invertible linear operator on $\CK$ for all 
$p\in H_{m}^+$ whose matrix elements, together with those of 
$L(p)^{-1}$, are polynomials in $p$. In fact, $L(\I\partial)$ is then 
an invertible matrix of differential operators (at least when operating in 
fields satisfying the Klein-Gordon equation) and the two fields are 
related by
\begin{equation}
\Phi_{\alpha}^{(1)}(x)=\hbox{$\sum_{\beta}$}L_{\alpha\beta}(\I\partial)
\Phi_{\beta}^{(2)}(x),\quad
\Phi_{\beta}^{(2)}(x)=\hbox{$\sum_{\alpha}$}L^{-1}_{\beta\alpha}(\I\partial)
\Phi_{\alpha}^{(1)}(x).
\end{equation}
For a true counterexample to PCT symmetry of the QFTLO generated by a
free field $\Phi_{\alpha}^{(1)}$ one must therefore pick  $M^{(1)}(p)$ 
in such a way that it can {\it not} be written as \eqref{equalalgebras} 
with a PCT symmetric $ M^{(2)}(p)$. This, however can easily be 
shown e.g.\ for $M$ defined by the matrix
\begin{equation} M(p)=\left(\begin{array}{cc}
(p^0)^2  & \I m p^0\\
 - \I m p^0 & (p^0)^2
\end{array} \right). \end{equation}
It should even be possible to find within the class of Lorentz non-covariant, 
finite component free 
fields examples satisfying wedge 
duality, but still violating PCT symmetry. In fact, it is easy to check 
that wedge duality certainly 
holds, if $M(p)^{-1}$ exists for all for all 
$p\in H_{m}^+$ and has polynomial matrix elements. A sufficient, and 
by Lemma V.2 in \cite{GY00} also necessary, condition for this is that 
$\det M(p)$ is constant on the mass shell. PCT symmetry, on the other 
hand, requires that $M(p)$ can be written as $L(p)^{*} 
M^{\prime}(p)L(p)$ with $M^{\prime}(p)$ satisfying 
\eqref{pctsymmfield} and $L(p)$ polynomial in $p$. Although an
explicit example 
satisfying wedge duality without the latter property is not known to 
us there is hardly a doubt that such examples exist. 

If one allows infinite dimensional index spaces $\CK$ Oksak and Todorov 
\cite{OT68} have given examples of free fields violating PCT symmetry, but 
with Lorentz covariance, i.e., where there is a representation  
$\Lambda\to V(\Lambda)$ of the Lorentz group on $\CK$ such that
\begin{equation}
    M_{\sigma,\rho}(\Lambda p)= 
M_{V(\Lambda^{-1})\sigma,V(\Lambda^{-1})\rho}(p).
    \label{eq:covar}
\end{equation}
All examples with infinite dimensional $\CK$ violate the split 
property, but the status of wedge duality in this example, i.e., the 
validity of the analyticity 
condition of Theorem 2.6, is not known. 
The full analyticity condition of Theorem 2.15 is 
certainly violated.

Another instructive example is the case of an infinite number of 
copies of the neutral scalar free field. Here $\CK$ is infinite 
dimensional, but $M_{\sigma,\rho}(p)$ is independent of $p$: 
\begin{equation}
   M_{\sigma^*,\rho}(
    p)=\langle\sigma,\rho\rangle
    \label{}
\end{equation}
where $\langle\sigma,\rho\rangle$ is some scalar product on $\CK$. This 
theory is clearly invariant under the Poincar\'e transformations that 
simply ignore the index space $\CK$:
\begin{equation}
    U_1(a,\Lambda)\Phi(f,\sigma)U_1^*(a,\Lambda)=\Phi(
    f_{\{a,\Lambda\}},\sigma)
\end{equation}
with $f_{\{a,\Lambda\}}(x)=f(\Lambda^{{-1}}(x-a))$. This theory is also 
covariant under other representations of the Poincar\'e
group. Let $V(\Lambda)$ be a continuous unitary representation
of the Lorentz group acting on $\CK$. Then we may define a 
representation $U_{2}$ by $U_{2}(a,\Lambda)\Omega=\Omega$ and
\begin{equation}
 U_2(a,\Lambda)\Phi(f,\sigma)U_2^*(a,\Lambda)
    =\Phi(f_{\{a,\Lambda\}},V(\Lambda^{-1})\sigma)
\end{equation}
In this situation the cocycle linking $U_{1}$ and $U_{2}$ 
is itself a group representation and
$U_1$ is the minimal representation. This follows from the fact
that $U_1$ acts only on the test-functions which implies that
the Wightman functions fulfill the Hall--Wightman analyticity [HW57] on the
complex Lorentz group.

This is an example with wedge duality and the existence of two 
representations of the Poincar\'e group. One of them is the minimal 
one so that the Bisogano Wichmann property holds.
For the second representation one has the partial
analyticity property for $U_2(\Lambda_W(t))$ required for wedge duality, 
but not the full analyticity
property required for Theorem 2.15. The analyticity for a fixed wedge $W$ 
holds because $V(\Lambda_W(t))$ is a one-parametric
group, which has sufficiently many analytic elements. On the other hand, since 
$V(\Lambda)$ is a non-trivial unitary representation of the Lorentz
group it does not have an analytic continuation onto the complex
Lorentz group and therefore there are no fully analytic elements.

Finally we mention an example given in \cite{BDFS00}, Section 5.3, 
where the Bisognano-Wichmann property 
is violated, but the conditions of Theorem 2.18 are fulfilled so that 
Poinca\'e 
covariance (without spectrum condition) and PCT symmetry hold. 
In this example the split property is fulfilled. One may 
ask whether Poincar\'e covariance, spectrum condition and nuclearity
(which implies the split property) are sufficient to derive 
the PCT theorem. No counterexamples are known and 
Theorems 2.12 and 2.13 may be regarded as a step in this
direction, but this question is otherwise open.


\begin{thebibliography}{99}

%
\bibitem {BW75} Bisognano, J. and Wichmann, E.H.,
\textit{ On the duality condition for a Hermitian scalar field},
J. Math. Phys. \textbf{16} (1975), 985-1007.
%
\bibitem {BW76} Bisognano, J. and Wichmann,E.H.,
\textit{ On the duality condition for quantum fields},
J. Math. Phys. \textbf{17} (1976), 303-321.
%
\bibitem{Bch65} Borchers, H.J., 
\textit{ Local Rings and the Connection of Spin with Statistics},
Commun. Math. Phys. \textbf{1} (1965), 281-307.
%
\bibitem{Bch70} Borchers, H.J., 
\textit{On Groups of Automorphisms with Semi-bounded Spectrum}
In: Syst\`emes \`a un Nombre Infini de Degr\'es de Libert\'e, p. 125-142
Editions du C.N.R.S. Paris (1970).
%
\bibitem {Bch92} Borchers, H.-J.,
\textit{ The CPT-Theorem in Two-dimensional Theories of Local Observables},
Commun. Math. Phys. \textbf{143} (1992), 315-332.
%
\bibitem {Bch95} Borchers, H.-J., 
\textit{ When does Lorentz Invariance imply Wedge--Duality}, 
Lett. Math. Phys. \textbf{35} (1995), 39-60.
%
\bibitem{Bch96} {Borchers, H.-J.,  \textit{ Translation Group and
Particle Representations in Quantum Field Theory},
Lecture Notes in Physics \textbf{m40} Springer, Heidelberg, 1996.}
%
\bibitem{Bch96b} {Borchers, H.-J., \textit{ Half--sided Modular Inclusion
and the construction of the Poincar\'e group}, Commun. Math. Phys.
\textbf{179} (1996), 703-723.}
%
\bibitem {Bch98} {Borchers, H.-J., 
\textit{ On Poincar\'e transformations and the modular
group of the algebra associated with a wedge},
Lett. Math. Phys. \textbf{46}, (1998) 295-301.}
%
\bibitem {BY92} {Borchers, H.-J. and Yngvason, J.,
\textit{ From quantum fields to local von neumann algebras},
Rev. Math. Phys. \textbf{Special issue}, (1992) 15-47.}
%
\bibitem {BGL93} {Brunetti, R., Guido,D. and Longo, R.,
\textit{ Modular Structure and Duality in Conformal Quantum
Field Theory}, Commun. Math. Phys. \textbf{156} (1993), 201-219.}
%
\bibitem {BGL94} {Brunetti, R., Guido, D. and Longo, R.,  
{ Group cohomology, modular theory and space--time symmetries},
Rev. Math. Phys. \textbf{7} (1994), 57-71.}
%
\bibitem {BDF87} {Buchholz, D., D'Antoni, C. and Fredenhagen, K.,
\textit{ The universal structure of local algebras},
Commun. Math. Phys. \textbf{84} (1987), 123-135.}
%
\bibitem {BS93} {Buchholz, D. and Summers, S.J.,
\textit{ An Algebraic Characterization of Vacuum States in Minkowski Space},
Commun. Math. Phys. \textbf{155} (1993), 449-458.}
%
\bibitem {BDFS00} {Buchholz, D., Dreyer, O., Florig, M. and Summers, S.,
\textit{ Geometric Modular Action and Spacetime Symmetry Groups},
Rev. Math. Phys. \textbf{4} (2000), 475-560.} 
%
\bibitem {BFS99} {Buchholz, D., Florig, M., and Summers, S., \textit{ An
Algebraic Characterization of Vacuum States in Minkowski Space},II,
Preprint (1999).}
%
\bibitem {BF82} {Buchholz, D. and Fredenhagen, K.,\textit{ Locality and
the structure of particle states}, Commun. Math. Phys. \textbf{84} (1982),
1-54.}
%
\bibitem{BuW86} {Buchholz, D. and Wichmann, E.H., \textit{ Causal independence
and energy level density of states in local quantum field theory},
Commun. Math. Phys. \textbf{106} (1986), 321.}
%
\bibitem {DHR69a} {Doplicher, S., Haag, R., and Roberts, J.E.,
\textit{ Fields, observables and gauge transformations} I,
Commun. Math. Phys. \textbf{13} (1969a), 1.}
%
\bibitem {DHR69b} {Doplicher, S., Haag, R., and Roberts, J.E.,
\textit{ Fields, observables and gauge transformations} II,
Commun. Math. Phys. \textbf{15} (1969a), 173.}
%
\bibitem{DL84} {Doplicher, S. and Longo, R., \textit{ Standard and split
inclusions of von Neumann algebras}, Invent. Math. \textbf{75} (1984), 
493-536.}
%
\bibitem{DR90} {Doplicher, S. and Roberts, J.E., \textit{ Why there is 
a field algebra with a compact gauge group describing the 
superselection structure of particle physics}, Commun. Math. Phys. 
\textbf{131} (1990), 187.}
%
\bibitem {Ep66} {Epstein, H.,
\textit{ Some Analytic Properties of Scattering Amplitudes in Quantum
Field Theory}, in 1965 Brandeis Summer Institute, Gordon and Breach,
New York, London, Paris, 1966.}
%
\bibitem{Florig} {Florig, M., \textit{On Borchers' theorem}, lett.\ 
math.\ Phys.\ {bf 46},(1998)  289--293.}

\bibitem {GY00} {Gaier, J., Yngvason, J., \textit{ Geometric Modular Action,
Wedge Duality and Lorentz Covariance are Equivalent for
Generalized Free Fields}, J. Math. Phys. {bf 41} (2000), 5910-5919.}
%
\bibitem {Gr61} {Greenberg, O.W., 
\textit{ Generalized Free Fields and Models of local Field Theory},
Ann. Phys. \textbf{16} (1961), 158-176.}
%
\bibitem{Gui95} {Guido, D., \textit{ Modular covariance, PCT, spin
and statistics}, Ann.\ Inst.\ H.\ Poincar\'e \textbf{64} (1995), 383-398.}
%
\bibitem {GL92} {Guido, D. and Longo,R., 
\textit{ Relativistic Invariance and Charge Conjugation in Quantum 
Field Theory}, Commun. Math. Phys. \textbf{148} (1992), 521--551.} 
%
\bibitem {GL95} {Guido, D. and Longo, R., \textit{ An Algebraic Spin and
Statistic Theorem}, Commun. Math. Phys. \textbf{172} (1995), 517-533.}
%
\bibitem {GL00} {Guido, D. and Longo, R., \textit{Natural Energy Bounds
in Quantum Thermodynamics}, Preprint (2000)}
%
\bibitem{Ha92} {Haag, R.,\textit{ Local Quantum Physics}, 
Springer Verlag, Berlin, Heidelberg, New York, 1992.}
%
\bibitem {HW57} {Hall, D. and Wightman, A.S., \textit{ A theorem on invariant
analytic functions with applications to relativistic quantum field theory},
Danske Vidensk. Selskab, Mat.-fysiske Meddelelser
\textbf{31,no 5} (1957), 1-41.}
%
\bibitem {Jo57} {Jost, R.,
\textit{ Eine Bemerkung zum CTP Theorem},
Helv. Phys. Acta \textbf{30} (1957), 409-416.}
%
\bibitem {Ku97} {Kuckert, B., \textit{ Spin \& statistics, localization
regions, and modular symmetries in quantum field theory},
PhD-thesis, Univ.\ Hamburg, 1997.}
%
\bibitem {Ku00} {Kuckert, B., \textit{ Two uniqueness results on the
Unruh effect and on the PCT-symmetry}, Preprint (2000).}
%
\bibitem {KW99} {K\"ahler, R. and Wiesbrock, H.-W.,
\textit{ Modular theory and the reconstruction of the 4-dimensional
quantum field theory}, Preprint (1999).}
%
\bibitem {Lue54} {L\"uders, G.,
\textit{ On the equivalence of invariance under time reversal and under
particle-antiparticle conjugation for relativistic field theories},
Danske Vidensk. Selskab, Mat.-fysiske Meddelelser
\textbf{28, no 5} (1954), 1-17.}
%
\bibitem {LZ58} {L\"uders, G. and Zumino, B., 
\textit{ On the connection between spin and statistics},
Phys. Rev. \textbf{110} (1958), 1450-1453.}
%
\bibitem{Mu00} {Mund, J.,\textit{ The Bisognano-Wichmann Theorem for
Massive Theories}, Preprint (2000).}  
%
\bibitem {OT68} {Oksak, A.I. and Todorov, I.T., \textit{ Invalidity of the 
TCP--Theorem for Infinite--Component Fields},
Commun. Math. Phys. \textbf{11} (1968), 125.}
%
\bibitem {Pau55} {Pauli, W.,
\textit{ Exclusion Principle, Lorentz Group and Reflection of Space-Time
and Charge}, in:
Niels Bohr and the Development of Physics,
W. Pauli, L. Rosenfeld, and V. Weisskopf ed., Pergamon Press, London, 1955.}
%
\bibitem{RS61} {Reeh, H. and Schlieder, S., \textit{ Eine Bemerkung zur 
Unit\"ar\"aquivalenz von Lorentzinvarianten Feldern}, 
Nuovo Cimento \textbf{22} (1961), 1051-1068.}
%
\bibitem {Str67} {Streater, R., \textit{ Local Fields with the Wrong Connection
Between Spin and Statistics}, Commun. Math. Phys. \textbf{5} (1967), 88-96.}
%
\bibitem{Wie97} {Wiesbrock, H.-W., \textit{ Symmetries and modular
intersections of von Neumann algebras}, Lett.Math.Phys. \textbf{39} (1997),
203-212.}
%
\bibitem{Wie98} {Wiesbrock, H.-W., \textit{ Modular
intersections of von Neumann algebras in quantum field theory},\newline 
Lett.Math.Phys. \textbf{39} (1998), 203-212.}
%
\bibitem{Wi56} {Wightman, A.S.,
\textit{ Quantum field theory in terms of vacuum expectation values},
Phys. Rev. \textbf{101} (1956), 860.}
%
\bibitem {Yng94} {Yngvason, J., \textit{ A Note on Essential Duality},
Lett.Math.Phys. \textbf{31} (1994), 127-141.}
%

\end{thebibliography}

\end{document}